\documentclass[preprint]{ptephy_v1}
\preprintnumber{2312.16029}

\usepackage{physics,amsmath,amssymb,mathrsfs,bm,bbm,indentfirst,color,physics,graphics,comment,marvosym,graphicx,hyperref}
\newcommand{\rev}[1]{#1}
\renewcommand{\grad}{\nabla}

\title{%
Dependence of the critical temperature and disorder in holographic superconductors on superfluid density
}
\author{
Zhenguo Wang$^{a}$\thanks{%
\href{mailto:wang_zg@shu.edu.cn}{\texttt{wang\_zg@shu.edu.cn}}
}
}
\author{
Xian-Hui Ge$^{a,b}$\thanks{%
\href{mailto:gexh@shu.edu.cn}{\texttt{gexh@shu.edu.cn} (corresponding author)}
}
}
\author{
Shuta Ishigaki$^{a}$\thanks{%
\href{mailto:shutaishigaki@shu.edu.cn}{\texttt{shutaishigaki@shu.edu.cn}}
}
}
\affil{$^{a}$\emph{Department of Physics, Shanghai University,
99 Shangda Road, Baoshan District,
Shanghai 200444, China}}
\affil{$^{b}$\emph{Shanghai Key Laboratory of High Temperature Superconductors,
Shanghai University,
99 Shangda Road, Baoshan District,
Shanghai 200444, China}}
\date{}

\begin{document}
\begin{abstract}
Recent experiments strongly indicate deep connections between transports of strange metal and high $T_c$ superconductors.
For instance, it is known that the dependence of the zero-temperature phase stiffness on the critical superconducting temperature becomes linear in underdoped materials.
In this paper, we investigate relation meticulously between the phase stiffness and the critical superconducting temperature for the Gubser-Rocha holographic superconductor model in the probe limit.
The superfluid density (or phase stiffness) can be extracted from the low-frequency dependence of the AC conductivity.
\rev{More importantly, we find linear dependence of zero-temperature superfluid density on the critical superconducting temperature, which has a similarity to a recent experiment in a film cuprate.}
In addition, we also provide some approximate formulas for the critical temperatures and the AC conductivity.
\end{abstract}

\maketitle
\newpage

\tableofcontents

\section{Introduction}
Superconductivity is a very significant discovery in physics in the early part of the 20th century.
Since its ascertainment, overwhelming experimental and theoretical research has been done on superconductivity, one of the most striking is Bardeen-Cooper-Schrieffer (BCS) theory\cite{bcs}.
This theory assumes that the fundamental characteristics of superconductivity are caused by Cooper's pair correlation.
As experiments progressed, high-temperature oxide superconductors\cite{muller,mkwu,hiroshi,zzsheng} and iron-based superconductors\cite{jjhamlin,kamihara,xhchen,haihuwen} were found.
The description of these superconductors is beyond the scope of BCS theory.
Recently, a series of groundbreaking experimental findings have emerged that demonstrate how the superfluid density increasingly manifests a linear relationship with temperature as temperatures escalate \cite{mm,soumyajit,sudhansu}. These phenomena depart from the conventional expectations set by the BCS theory, thereby shedding light on potentially novel facts of high-temperature superconductivity and its fundamental physics.
There are lots of interests and challenges in studying the critical temperature $T_c$ related to the superfluid density $n_s$.
Some experimental papers \cite{lulian,copperoxides,microwavetl2201} elucidate that $T_c$ seems to be principally controlled by the superfluid density. More remarkably, the superfluid density displays a strong linear temperature dependence within a certain temperature range.
In Ref.~\cite{copperoxides}, the authors have been reported the phase stiffness depend on temperature and doping by investigating the entire overdoped side of the $\text{La}_{2-x}\text{Sr}_{x}\text{CuO}_{4}$ (LSCO) phase diagram.
In their results, the dependence of the zero-temperature phase stiffness on the critical is mostly linear, but with an offset.
However, close to the origin this dependence becomes parabolic.
This phenomenon was taken as evidence for the violation of the Homes' law.
In Ref.~\cite{dordevic}, the authors have unraveled explicitly that the Homes' law is indeed violated, and as the main reason for violation is that the superfluid density in LSCO films is suppressed more strongly than in bulk samples. 

Based on the above experimental findings, in this paper, we explore the universal properties of the superconductors, e.g., scaling relations, specially concentrate on the dependence of the zero-temperature phase stiffness on the critical superconducting temperature, by using the AdS/CFT correspondence in a specific holographic model.
The AdS/CFT correspondence\cite{AdSCFT,Gubser:1998bc,Witten:1998qj}, also known as holography or gauge/gravity duality, is a framework for tracing the behavior of strongly-coupled quantum systems in terms of a higher dimensional gravity theory.
It can be a very powerful and fascinating tool to tackle strongly interacting systems by studying gravitational models.
In particular, holographic techniques have been widely applied for studying condensed matter physics \cite{zaanen,Baggioli:2019rrs,ammon,Cai:2015cya}.

The kind of model called holographic superconductors is still an attractive topic of theoretical research.
They have the potential to provide new insights into the nature of superconducting materials and could give rise to the development of new technologies based on superconductivity. 
Based on the AdS/CFT correspondence, the first minimal bottom-up construction had been implemented by Hartnoll, Herzog, and Horowitz\cite{bhs}.
In Ref.~\cite{bhs}, the authors considered the charged complex scalar and the Maxwell field in the Schwarzschild-AdS$_{4}$ (SAdS$_4$) spacetime.
Although this is a simple model, it can reproduce the typical behavior of the superconductor successfully.
The calculation done in \cite{bhs} was the so-called probe limit, ignoring the backreactions to the gravity sector from the matter sector.
In \cite{Hartnoll:2008kx}, the authors considered the Einstein-Maxwell theory with the complex scalar and found that the probe-limit calculation corresponds to the infinite charge case with appropriate scaling.
In any case, the complete analysis of the superconducting phase requires numerical computations.
Meanwhile, some analytic properties of the holographic superconductor in the superconducting phase have been studied\cite{Herzog:2009xv,Gary,Siopsis:2010uq,Ge:2010aa}.
Since the complete analysis requires numerical computations, they provided some approximate formulas.
From then on \cite{bhs,Hartnoll:2008kx}, many papers have appeared that attempt to apply this useful idea to investigate superconductors in different setups \cite{Franco:2009yz,Gangopadhyay:2013qza,Cai:2010cv,Roychowdhury:2012hp,Sheykhi:2016kqh,Qiao:2020hkx,Erdmenger:2013zaa,Ge:2015fmu,Zhao:2023qms}.

Recently, it has been recognized that the strange metallic behavior is universal in superconductors like the high-$T_c$ cuprates.
The strange metal is often described by the linear-$T$ dependence of its resistivity\cite{yuan,jiang,Phillips:2022nxs}.
In holographic studies, the Gubser-Rocha model\cite{Gubser:2009qt} is known as one of the candidates describing the strange metallic behavior of the superconductors.
This model is an Einstein-Maxwell-dilaton theory, and axion fields are often introduced to break the translational invariance\cite{Andrade:2013gsa,Gouteraux:2014hca}.
The original model is consistent with the specific truncation of the 11-dimensional supergravity.
We can explore microscopic mechanism within this model such as quantum critical points and superconducting transition temperature, which help us understand the origin of high-temperature superconductivity more profoundly.
This model has been widely utilized for studying the holographic version of the high-$T_c$ cuprates superconductors. 
As well known, the Homes' law is a universal relation of superconductors between the superfluid density $n_s$ at zero temperature, the cirtical temperature $T_c$ and the direct current (DC) conductivity $\sigma_{DC}$, which is observed in high $T_c$ superconductors.
In Ref.~\cite{Homes' law}, the authors have investigated how Home's law can be realized together with linear-$T$ resistivity by using holographic models related to the Gubser-Rocha model.
\rev{
Inspired by literature \cite{copperoxides,dordevic}, our main goal is to study the phenomena that deviate from Holmes' law.
Because there has been no study for such phenomenon with alternating current (AC) conductivity in holography, our study is a important and essential step to look into the novel experimental results.
}
On the other hand, we notably advance the analytical framework for studying holographic superconductors by developing novel approximate methods to explore key analytic properties within the context of the Gubser-Rocha model. 
Despite the prevalent reliance on numerical computations in previous studies, akin to those conducted in the foundational holographic superconductor paradigm, we focus our efforts here on investigating analytically significant aspects such as the critical temperature and AC conductivity.

The primary objective of this paper is to investigate the scaling relationships connecting the superconducting critical temperature, the superfluid density, and the momentum-dissipation strength within the framework of the holographic superconductor model derived from the Gubser-Rocha model. Moreover, a significant part of our study involves comparing these theoretically deduced relations with empirical results obtained in experimental settings.
First, we study the critical temperature by finding the onset of the charged-scalar instability.
Taking advantage of the Sturm-Liouville method, we investigate how the momentum-dissipation strength affects the critical temperature.
The results obtained by numerical and approximate methods are in good agreement with each other.
Using the approximate formula, we find the scaling relation between the critical temperature and the strength of the momentum dissipation.
Subsequently, we compute the AC conductivity in the superconducting state.
To extract the superfluid density, we meticulously analyze the AC conductivity data and proceed to examine the correlation between the zero-temperature superfluid density and the critical temperature.
\rev{
Most notably, our results roughly show a linear relationship between the zero-temperature phase stiffness and the critical temperature.
This result is similar to recent experimental outcomes observed in a certain overdoped copper-oxide film \cite{copperoxides}.
Interestingly, this material does not obey the Homes' law \cite{dordevic}.
}

The organization of this paper is as follows. 
The holographic setup of our model is given in section \ref{setup}. 
We investigate the critical temperature of the superconducting phase in section \ref{section3}. 
In section \ref{section4}, we turn to study the AC conductivity for obtaining the superfluid density.
We provide conclusion and discussion in section \ref{section5}.
In Appendix \ref{appendixa}, we demonstrate how the profile of the critical temperature with general $q$ from the full analysis with backreactions agrees with the probe limit results in a large $q$ limit.
In Appendix \ref{appendix:low-T}, we show some results of the approximate formula for the AC conductivity in the ordered phase.

\section{Setup}\label{setup}
In this study, we employ the Gubser-Rocha model \cite{Gubser:2009qt} extended with the linear axions and the charged scalar, for studying the superconducting nature \cite{Zhou:2015qui, Kim:2017dgz, Jeong:2018tua, Homes' law}.
Particularly, we study the relation between superfluid density $n_s$ and superconducting critical temperature $T_c$ which has not been explored so far.
The action of our model is given by \cite{Jeong:2018tua,Homes' law}
\begin{subequations}\label{eq:full_model}
\begin{align}
	S =& \int d^4 x\sqrt{-g} \left( \mathcal{L}_{g} + \mathcal{L}_{m} \right),\\
	\mathcal{L}_{g} =&
	R - \frac{1}{2} (\partial \phi)^2  + 6 \cosh \frac{\phi}{\sqrt{3}} - \frac{1}{2} \sum (\partial \psi_I)^2,\\
	\mathcal{L}_{m} =&
	- \frac{1}{4} e^{\frac{\phi}{\sqrt{3}}} F^2
	- \abs{D\Phi}^2 - B(\phi) \abs{\Phi}^2.
\end{align}
\end{subequations}
The action consists of two components, $\mathcal{L}_{g}$ representing the Lagrangian of fields involving the gravity and $\mathcal{L}_{m}$ representing the Lagrangian of matter fields.
The model involves gravity fields $g_{\mu\nu}$, a dilaton $\phi$, axions $\psi_{I}$, $U(1)$ gauge fields and a complex scalar field $\Phi$.
It was shown that the presence of the dilaton field can realize the vanishing entropy at zero temperature\cite{Davison:2013txa}.
The axions are originally introduced to break the translational invariance explicitly in the full analysis with backreactions \cite{Zhou:2015qui}.
$F_{\mu\nu}= \partial_{\mu} A_{\nu} - \partial_{\nu} A_{\mu}$ is the field strength of the $U(1)$ gauge field.
The covariant derivative is defined by $D_{\mu}:=\grad_{\mu}-iqA_{\mu}$.
$B(\phi)$ is a coupling factor between dilaton field $\phi$ and complex scalar field $\Phi$.

We concentrate our analysis in the probe limit, which corresponds to taking the limit as q approaches infinity. Under this particular condition, we can confidently assert that the gravity   $\mathcal{L}_{g}$ and matter sectors $\mathcal{L}_{m}$ are effectively decoupled and described independently from one another. The equations of motion for the dilaton and the axions are obtained from $\mathcal{L}_g$ as
\begin{equation}\label{eq:eom_dilaton_axions}
	\grad^{2}{\phi}
 +2\sqrt{3}\ \text{sinh}\left(\frac{\phi}{\sqrt{3}}\right)=0,\quad
    \grad^{2}{\psi_{I}}=0,
\end{equation}
and the Einstein's equation becomes
\begin{align}\label{eq:Einstein-eq}
	R_{\mu\nu}-\frac{1}{2}g_{\mu\nu}\left[R
    -\frac{1}{2}(\partial \phi)^{2}
    +6\cosh\left(\frac{\phi}{\sqrt{3}}\right)-\frac{1}{2}\sum_{I=1}^{2}\left(\partial \psi_{I}\right)^{2}
    \right]
	=
    \frac{1}{2}\partial_{\mu}\phi\partial_{\nu}\phi+\frac{1}{2}\sum_{I=1}^{2}\left(\partial_{\mu}\psi_{I}\partial_{\nu}\psi_{I}\right).
\end{align}
Considering the metric ansatz
\begin{equation}\label{eq:metric}
	ds^{2}=-f(r)dt^{2}+\frac{dr^{2}}{f(r)}+h(r)\left(dx^{2}+dy^{2}\right), 
\end{equation}
we obtain the following solution \cite{Zhou:2015qui}
\begin{subequations}\label{eq:background_geometry}
\begin{equation}
	f(r)=r^{2}\left(1-\frac{P}{r}\right)^{1/2}\left(1-\frac{k^{2}}{2r^{2}}\right),\quad
    h(r)=r^{2}\left(1-\frac{P}{r}\right)^{1/2},
\end{equation}
\begin{equation}
	\phi(r)=-\frac{\sqrt{3}}{2}\ln\left(1-\frac{P}{r}\right),\quad
    \psi_I = (k x, k y),
\end{equation}
\end{subequations}
where $P$ is an integration constant corresponding to the location of the curvature singularity at $r=P$ and it can be expressed in terms of temperature $T$ and momentum dissipation strength $k$, see Eq.~\eqref{eq:temperature}.%
\footnote{
    Note that the blackening factor is still different from $f(r) = r^2(1-r_h^3/r^3)$ in the SAdS$_4$, even if one sets $P=0$.
}
$k$ is a positive constant of the linear axions, which can be understood as the strength of the momentum dissipation in the backreacted setup.
Remark that $k$ was introduced to break the translational symmetry.
In this study, we naively expect that $k$ still holds some aspects of the strength of the momentum dissipation, even in the probe limit.
The geometry is a black hole spacetime with the horizon at $r=r_h = k/\sqrt{2}$.
To avoid the naked singularity, the range of $P$ is restricted by $-\infty<P<r_h$.
In the probe limit, the gravity sector is decoupled from the Maxwell field.
Therefore, this solution is a neutral black hole solution.
Note that this solution was also studied in Refs.~\cite{Ren:2021rhx,Jeong:2023ynk}, and Ref.~\cite{Ren:2022qkr} without the axions.
This is also a special neutral case of the charged black hole studied in Refs.~\cite{Zhou:2015qui,Caldarelli:2016nni,Kim:2017dgz,Jeong:2018tua, Homes' law}.
(See also Appendix \ref{appendixa}).
The authors of Ref.~\cite{Homes' law} studied the same model for general $q$ beyond the probe limit.
We call the above solution as `dilatonic black hole'.
The Hawking temperature and the entropy density read
\begin{equation}\label{eq:temperature}
    T = \frac{r_h}{2\pi}\sqrt{1-\frac{P}{r_h}},\quad
    s = 4\pi r_{h}^2 \sqrt{1- \frac{P}{r_{h}}},
\end{equation}
respectively.
In order to obtain positive $T$, the range of $P$ is restricted in $-\infty<P<r_h$.
We can write the entropy density $s=4\sqrt{2}\pi^2 k T$.

On the other hand, the model also admits another neutral solution without the dilaton hair, which is given by \cite{Andrade:2013gsa}
\begin{equation}\label{eq:bald_black_hole}
    f(r) = r^2 \left[
        1 - \frac{k^2}{2 r^2} - \frac{r_{h}^3}{r^3}\left(
            1 - \frac{k^2}{2 r_{h}^2}
        \right)
    \right],\quad
    h(r) = r^2,\quad
    \phi(r) = 0,\quad
    \psi_{I} = (k x, k y).
\end{equation}
We call it as `bald black hole' solution in this study.
In this solution, $k$ and $r_{h}$ are independent parameters but the number of parameters does not change.
Since the dilaton vanishes, the solution is same as the neutral solution in the linear axions model \cite{Andrade:2013gsa}.
Now, the Hawking temperature and the entropy density read
\begin{equation}\label{eq:T_bald}
    T = \frac{1}{8\pi} \frac{6 r_{h}^2 - k^2}{r_{h}},\quad
    s = 4\pi r_{h}^2,
\end{equation}
respectively.
The range of $k$ is restricted as $k<\sqrt{6} r_h$ by $T>0$.
This solution is favored in the range $T > \frac{k}{2\sqrt{2}\pi}$ thermodynamically\cite{Kim:2017dgz,Jeong:2023ynk}.
Therefore, we need to consider this solution as the background metric rather than the dilatonic black hole (\ref{eq:background_geometry}), at higher temperature where $T > k/(2\sqrt{2}\pi)$.

Under the probe limit, we can analyze the matter sector by considering the gravity, dilaton, and axions as non-dynamical background fields given by Eq.~(\ref{eq:background_geometry}).
For the matter sector, the equations of motion are obtained as
\begin{equation}\label{maxwelleq}
	\grad_{\mu}{\left(e^{\frac{\phi}{\sqrt{3}}}F^{\mu\nu}\right)}-iq\Phi^{*}\left(\partial^{\nu}-iqA^{\nu}\right)\Phi+iq\Phi\left(\partial^{\nu}+iqA^{\nu}\right)\Phi^{*}=0,
\end{equation}
\begin{equation}\label{Phieq}
	D_{\mu}D^{\mu}\Phi-B(\phi)\Phi=0.
\end{equation}
We choose the mass term of $\Phi$ as
\begin{equation}\label{couplingfactor}
    B(\phi)=M^{2}\text{cosh}(\tau \phi),
\end{equation}
where $M$ is a mass of the scalar field near $r=\infty$, and $\tau$ is a constant determining the coupling with the dilaton field.
In this study, we fix $M^2=-2$, which leads the dimension of the charged-scalar operator as $\Delta = 1$ or $2$.
The authors of Ref.~\cite{Homes' law} indicate that the superconducting instability can be triggered easily at higher coupling $\tau$. 
For more detail argument about $B(\phi)$, see Ref.~\cite{Cremonini:2016bqw}.
Now, we consider the following ansatz:
\begin{equation}
    \Phi = \Phi(r),\quad
    A = A_t(r)\dd{t},
\end{equation}
where $\Phi(r)$ can be considered as a real function.
With the above ansatz, the equations of motion are written as
\begin{equation}\label{Phieqr}
	\Phi^{\prime\prime}(r)+\left(\frac{f^{\prime}(r)}{f(r)}+\frac{h^{\prime}(r)}{h(r)}\right)\Phi^{\prime}(r)+\frac{1}{f(r)}\left(\frac{q^{2}A_{t}(r)^{2}}{f(r)}-B(\phi)\right)\Phi(r)=0,
\end{equation}
\begin{equation}\label{Ateq}
	A_{t}^{\prime\prime}(r)+\left(\frac{h^{\prime}(r)}{h(r)}+\frac{\phi^{\prime}(r)}{\sqrt{3}}\right)A_{t}^{\prime}(r)-\frac{2e^{-\frac{\phi(r)}{\sqrt{3}}}q^{2}\Phi(r)^{2}}{f(r)}A_{t}(r)=0  \ .
\end{equation}
Near the boundary, these fields have the following asymptotic expansions
\begin{equation}\label{Phibehavior}
	\Phi(r)=\frac{\Phi^{(1)}/\sqrt{2}}{r}+\frac{\Phi^{(2)}/\sqrt{2}}{r^{2}}+\cdots \ ,
\end{equation}
\begin{equation}\label{At}
	A_{t}(r)=\mu-\frac{\rho}{r}+\cdots \ .
\end{equation}
For the vector field, $ \mu $ and $ \rho $ are dual to the chemical potential and charge density of the boundary conformal field theory, respectively.
For the scalar field, both the leading and the subleading terms can be normalizable in our case \cite{Witten:2001ua}.
Similarly to Ref.~\cite{bhs}, we refer the case when $\Phi^{(i)} = \expval{O_{i}}$ and $\Phi^{(j)} = 0$ as $O_{i}$ theory, where $i=1$ or $2$ and $j\neq i$.
The choice of the $O_{2}$ theory is called standard quantization, whereas the $O_{1}$ theory is realized by adding corresponding boundary action.
$\expval{O_i}$ is considered as an operator with dimension $i$ in the boundary theory.

For later convenience, we rewrite the coordinate by $z:=r_h/r$.
In this coordinate, the horizon is located at $z=1$, and the boundary at $z=0$.
Under this transformation, Eqs.~\eqref{Phieqr} and \eqref{Ateq} become
\begin{equation}\label{Phiz}
\Phi^{\prime\prime}(z)+\left(\frac{2}{z}+\frac{f^{\prime}(z)}{f(z)}+\frac{h^{\prime}(z)}{h(z)}\right)\Phi^{\prime}(z)+\frac{r_{h}^{2}}{z^{4}}\left(\frac{q^{2}A_{t}(z)^{2}}{f(z)^{2}}-\frac{B(\phi(z))}{f(z)}\right)\Phi(z)=0 \ ,
\end{equation}
\begin{equation}\label{Atz}
	A_{t}^{\prime\prime}(z)
    -\frac{2e^{\frac{-\phi(z)}{\sqrt{3}}}q^{2}r_{h}^{2}\Phi(z)^{2}}{z^{4}f(z)}A_{t}(z)=0 \ ,
\end{equation}
respectively.
Note that the prime denotes the derivative with respect to $z$ here.

\section{Critical temperature and condensation}\label{section3}
In this section, we investigate the phase boundaries by using both the numerical and analytic approximate methods.
Using the approximate result, we find the linearly scaling relation between $T_c$ and $k$ as \eqref{eq:scaling_Tc-k} for large $k$ in the $O_1$ theory.
We also present numerical results of the scalar condensation.

\subsection{Phase boundaries}
\rev{
In the normal phase, i.e., $\Phi=0$, Eq.~\eqref{Atz} becomes
\begin{equation}\label{Atcri}
	A_{t}^{\prime\prime}(z)=0 .
\end{equation}
With the boundary condition \eqref{At} and $A_{t}(r_h)=0$, the solution of this equation reads
\begin{equation}\label{eq:At_normal}
	A_{t}(z)=\mu(1-z),\qquad \mu=\frac{\rho}{r_{h}},
\end{equation}
where $\mu$ is an integration constant that can be read as the chemical potential in the boundary theory and
$\rho$ can be regarded as the charge density.
}

Below $T=T_c$, the normal phase solution will be unstable for the perturbation of the charged scalar.
We investigate the phase boundaries by finding the onset of the scalar instability.
In the $z$ coordinate, the expansion of $ \Phi $ is written as
\begin{equation}
	\Phi(z)=\frac{\Phi^{(1)}}{\sqrt{2}}\frac{z}{r_{h}}+\frac{\Phi^{(2)}}{\sqrt{2}}\frac{z^{2}}{r_{h}^{2}}+\cdots.
\end{equation}
As we have mentioned, we have two choices of the normalizable modes.
\rev{
We consider the following boundary conditions for each theory.
For the $O_{1}$ theory, we write $\Phi(z)$ as
\begin{equation}\label{O1}
	\Phi(z)=\frac{\langle O_{1}\rangle}{\sqrt{2}}\frac{z}{r_{h}}F(z),
    \quad F^{\prime}(0)=0,
\end{equation}
and the normalization is fixed by $F(0)=1$.
For the $O_{2}$ theory, we write $\Phi(z)$ as
\begin{equation}\label{O2}
	\Phi(z)=\frac{\langle O_{2}\rangle}{\sqrt{2}}\frac{z}{r_{h}^{2}}F(z), \quad F(0)=0,
\end{equation}
and the normalization is fixed by $F^{\prime}(0)=1$.
We also impose the regular condition for $ F(z) $ at $ z=1 $.
}\rev{
The problem can be understood as a Sturm-Liouville (SL) problem.
With Eq.~\eqref{eq:At_normal}, the equation of motion \eqref{Phiz} becomes
\begin{equation}
	\frac{d}{dz}\left\{p(z)F^{\prime}(z)\right\}+q(z)F(z)+\lambda r(z)F(z)=0,
\end{equation}
where
\begin{equation}\label{SLform}
\begin{split}
	p(z)&=z^{4}\tilde{f}(z)\tilde{h}(z),\\
	q(z)&=z^{3}\tilde{f}(z)\tilde{h}(z)\left(
        \frac{2}{z}+\frac{\tilde{f}'(z)}{\tilde{f}(z)}+\frac{\tilde{h}'(z)}{\tilde{h}(z)}
    \right)- B(\phi) \tilde{h}(z),\\
	r(z)&=\frac{(1-z)^{2}\tilde{h}(z)}{\tilde{f}(z)},\qquad
	\lambda =\frac{q^{2}\mu^{2}}{r_{h}^{2}}.
\end{split}
\end{equation}
We also defined $\tilde{f}(z) = r_{h}^{-2} f(z)$, and $\tilde{h}(z) = r_{h}^{-2} h(z)$.
}

\rev{
In order to obtain $T_{c}$, we need finding solution that is regular at the horizon, and satisfies Eq.~\eqref{O1} or \eqref{O2} at the AdS boundary, corresponding to the $O_{1}$ or $O_{2}$ theory.
This is an eigenvalue problem with an eigenvalue $\lambda$.
It can be solved numerically by using the shooting method.
In the case of fixing $\mu$, i.e., grand canonical ensemble, we write
\begin{equation}\label{eq:Tc-k}
    \frac{T_{c}}{q\mu}
    =
    \begin{cases}
    \frac{1}{2\pi}\sqrt{\frac{1-P/r_{h}}{\lambda_{c}}} & \text{(dilatonic)}\\
    \frac{6 - k^2/r_{h}^2}{8\pi \sqrt{\lambda_{c}}} & \text{(bald)}
    \end{cases},\quad
    \frac{k}{q \mu}
    =
    \begin{cases}
        \frac{\sqrt{2}}{\sqrt{\lambda_{c}}} & \text{(dilatonic)}\\
        \frac{k/r_{h}}{\sqrt{\lambda_{c}}} & \text{(bald)}
    \end{cases},
\end{equation}
by using \eqref{eq:temperature} and \eqref{eq:T_bald}.
The largest $T_{c}$ is usually related to the phase transition, so we find the smallest $\lambda_{c}$, correspondingly.
}

Moreover, we can also study the approximation for the lowest eigenvalue, as follows.
\rev{
A similar method was utilized in \cite{Siopsis:2010uq}.
}
Supposing that $F = \Psi_{n}$ is the eigenfunction for $ n $-th eigenvalue $ \lambda=\lambda_{n} $, we can formally write $ \lambda_{n} $ as
\begin{align}
	\lambda_{n}&=-\frac{\int_{0}^{1}\left[ \partial_{z}(p(z)\Psi_{n}^{\prime}(z))\Psi_{n}(z)+q(z)\Psi_{n}(z)^{2}\right]dz}{\int_{0}^{1} r(z)\Psi_{n}(z)^{2}dz}\nonumber\\
	&=\frac{\left.-p(z)\Psi_{n}(z)\Psi_{n}^{\prime}(z)\right|_{0}^{1}+\int_{0}^{1}\left[p(z)\Psi_{n}^{\prime}(z)^{2}-q(z)\Psi_{n}(z)^{2}\right]dz}{\int_{0}^{1} r(z)\Psi_{n}(z)^{2}dz}.
    \label{eq:RQ}
\end{align}
We leverage several properties of the Sturm-Liouville eigenvalue problem.
One of them is the set of eigenfunctions is complete.
\rev{
Thus, any regular function satisfying the boundary conditions can be expanded as $\sum_{n=1}^{\infty}c_{n}\Psi_{n}(z) $, where $c_{n}$ are coefficients of the expansion.
We can generalize this formula \eqref{eq:RQ} for a trial function, which is not an exact solution of the ordinary differential equation \eqref{SLform} but satisfies the boundary conditions.
In this study, we consider trial functions with one parameter $\alpha$ denoted by $F_{\alpha}(z)$.
We also define
\begin{equation}\label{lambdaF}
	\lambda_{(\alpha)} := \frac{\left.-p(z)F_{\alpha}(z)F_{\alpha}^{\prime}(z)\right|_{0}^{1}+\int_{0}^{1}\left[p(z)F_{\alpha}^{\prime}(z)^{2}-q(z)F_{\alpha}(z)^{2}\right]dz}{\int_{0}^{1} r(z)F_{\alpha}(z)^{2}dz},\quad
    \hat{\lambda}_{c} := \min_{\alpha} \lambda_{(\alpha)}.
\end{equation}
The function $\lambda_{(\alpha)}$ is bounded from below corresponding to that there is a minimum eigenvalue for the SL problem.
We can find the minimum value $\hat{\lambda}_{c}$, and it becomes an estimation for the actual smallest eigenvalue $\lambda_{c}$.
}

\begin{figure}
    \centering
    \includegraphics[width=8cm]{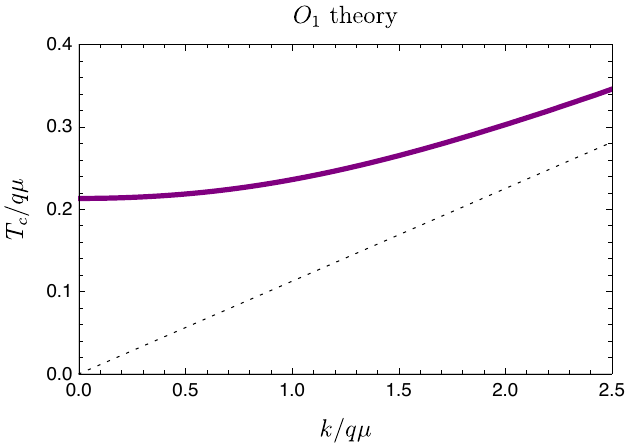}
    \includegraphics[width=8cm]{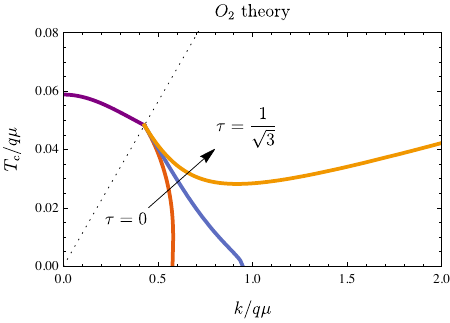}
    \caption{
        Correct superconducting phase boundaries for each theory.
        The black dotted lines show $T=k/(2\sqrt{2}\pi)$.
        The purple curves in $T>k/(2\sqrt{2}\pi)$ show the numerical results in the bald black hole.
        The other curves in $T<k/(2\sqrt{2}\pi)$ show the numerical results in the dilatonic black hole for various $\tau$.
        We have showed only the relevant results by taking the phase transition of the background geometry into account.
        In the right panel, the curves in $T<k/(2\sqrt{2}\pi)$ correspond to $\tau=0, \frac{3/4}{\sqrt{3}}, \frac{1}{\sqrt{3}}$ from bottom to top, respectively.
    }
    \label{fig:Tc-k_new}
\end{figure}

We remark that the background spacetime also exhibits the phase transition between Eq.~(\ref{eq:background_geometry}) and Eq.~(\ref{eq:bald_black_hole}).
Since we are working in the probe limit, this phase transition occurs at $T=k/(2\sqrt{2}\pi)$, regardless the charged-scalar instability.
In the following, we first explore the charged-scalar instability in each background geometry.
After that, we will show the correct phase boundaries as Fig.~\ref{fig:Tc-k_new} by combining the results obtained in each geometry.

\begin{figure}
    \centering
    \includegraphics[width=8cm]{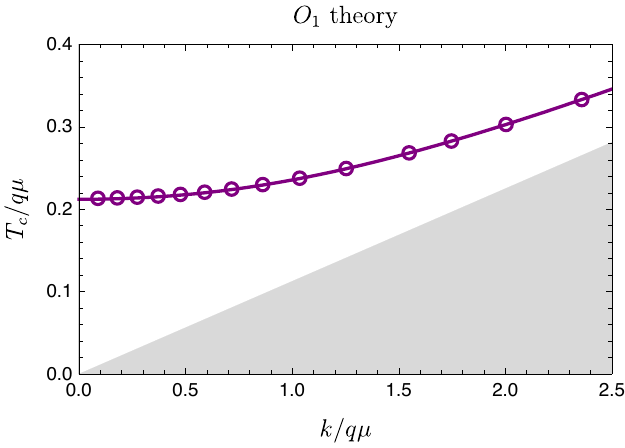}
    \includegraphics[width=8cm]{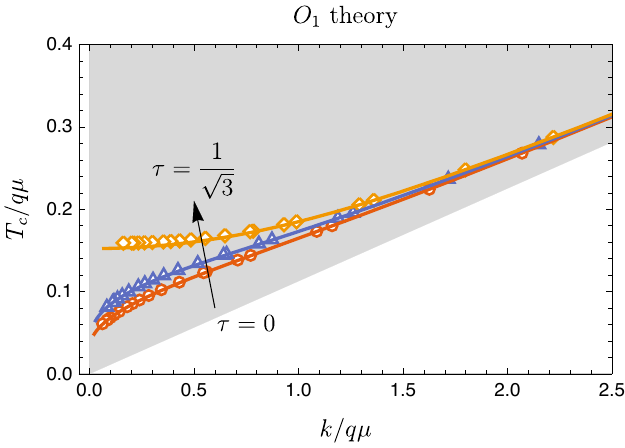}
    \caption{
    Superconducting phase boundaries in the $\frac{T_{c}}{q\mu}$--$\frac{k}{q\mu}$ plane for the $O_1$ theory in the two different black hole solutions.
    \rev{
    Each gray region indicates the thermodynamically unstable regime of the each background spacetime; the border is given by $T=k/(2\sqrt{2}\pi)$.
    }
    (left) The result in the bald black hole (\ref{eq:bald_black_hole}).
    (right) the result in the dilatonic black hole (\ref{eq:background_geometry}).
    The curves show the SL approximations while the points show the direct numerical results.
    In the right panel, the results depend on $\tau$.
    The curves and data correspond to $\tau = 0, \frac{3/4}{\sqrt{3}}, \frac{1}{\sqrt{3}}$ from bottom to top.
    }
    \label{fig:tc-kcompare}
\end{figure}

First, we consider the $O_1$ theory in each background geometry.
We choose the following trial function with one parameter $\alpha$ in both background black-hole geometry:
\begin{equation}\label{O1trialfunc}
	F_{\alpha}(z)=1+\alpha z^{2}.\quad (O_{1}\text{-theory})
\end{equation}
It satisfies the boundary conditions, $F(0)=1$ and $F^{\prime}(0)=0$.
\rev{
Substituting this into \eqref{lambdaF}, we obtain the approximate formula for the minimal eigenvalue, $\hat{\lambda}_c$.
Subsequently, we obtain $\frac{T_{c}}{q\mu}$ and $\frac{k}{q\mu}$ by using Eq.~\eqref{eq:Tc-k} with $\hat{\lambda}_c$.
The results are parameterized by $k/r_{h}$ in the bald black hole, and $P/r_{h}$ in the dilatonic black hole, respectively.
We illustrate phase boundaries by plotting these one-parameter curves in each background geometry.
}
Figure \ref{fig:tc-kcompare} shows the phase boundaries in the $\frac{T_c}{q\mu}$--$\frac{k}{q\mu}$ plane obtained by the approximation, and also by the numerical way for comparison.
Note that $q$ can be scaled out in the probe limit analysis.%
\footnote{
    In the backreacted case, $q$ can no longer be scaled out; different $q$ explains different setups.
    It is expected that the probe limit corresponds to $q\to\infty$ limit.
    See Appendix \ref{appendixa}.
}
The approximation agrees with the numerical result well.
In the $O_1$ theory, the phase boundaries always lie in on $T>k/(2\sqrt{2}\pi)$ in two background geometries, as shown in Fig.~\ref{fig:tc-kcompare}.
As a result, only the result in the bald black hole is relevant because the bald black hole is the ground state of the gravity sector when $T>k/(2\sqrt{2}\pi)$.

\begin{figure}[tbh]
	\centering
    \includegraphics[width=8cm]{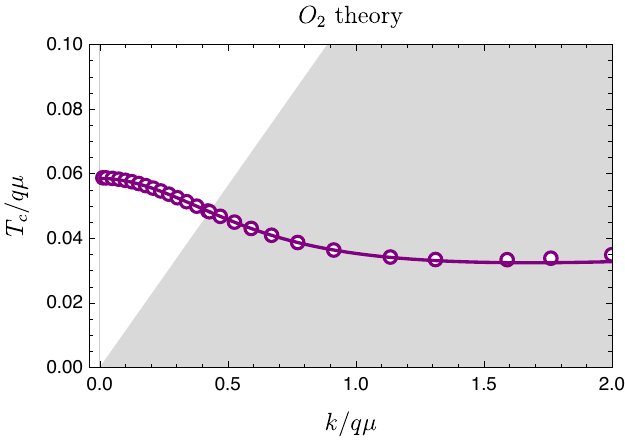}
    \includegraphics[width=8cm]{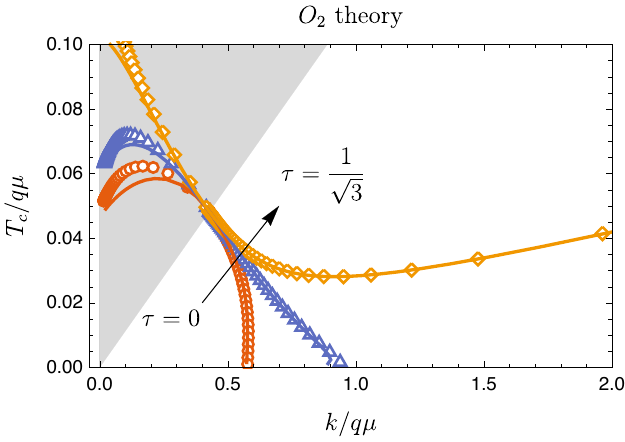}
	\caption{
    Superconducting phase boundaries in the $\frac{T_{c}}{q\mu}$--$\frac{k}{q\mu}$ plane for $O_2$-theory in the two different black hole solutions.
    \rev{
    Each gray region indicates the thermodynamically unstable regime of the each background spacetime; the border is given by $T=k/(2\sqrt{2}\pi)$.
    }
    (left) The result in the bald black hole (\ref{eq:bald_black_hole}).
    (right) The result in the dilatonic black hole (\ref{eq:background_geometry}).
    The curves show the SL approximations, while the points show numerical results obtained directly.
    In the right panel, the results depend on $\tau$.
    The curves and data correspond to $\tau = 0, \frac{3}{4\sqrt{3}}$ and $\frac{1}{\sqrt{3}}$ from bottom to top.
    }
	\label{fig:tc-kimprove}
\end{figure}

In the case of the bald black hole, the curve is parameterized by $k/r_{h}$ in $0<k/r_{h}<\sqrt{2}$.
In $k/r_{h} \to \sqrt{2}$, the curve approaches to $T=k/(2\sqrt{2}\pi)$.
In the case of the dilatonic black hole, the curves are parameterized by $P/r_{h}$ in $-\infty<P/r_h<0$, and small $|P|/r_h$ corresponds to large $\frac{k}{q\mu}$.
Expanding the approximation for $\tau=0$ in small $\epsilon \equiv - P/r_h$, we obtain
\begin{equation}
    \hat{\lambda}_{c} = \frac{1}{4 \ln 2 - 2} \epsilon + \order{\epsilon^2}.
\end{equation}
Using this leading expansion, $\frac{T_{c}}{q\mu}$ and $\frac{k}{q\mu}$ can be written as
\begin{equation}
    \frac{T_c}{q\mu}
    =
    \frac{\sqrt{\ln 2 - 1/2}}{\pi} \epsilon^{-1/2} + \order{\epsilon^{1/2}},\quad
    \frac{k}{q\mu}
    =
    2\sqrt{2\ln 2 - 1} \epsilon^{-1/2} + \order{\epsilon^{1/2}}.
\end{equation}
We obtain the scaling relation for large $k$ as
\begin{equation}\label{eq:scaling_Tc-k}
    T_c \approx \frac{k}{2\sqrt{2}\pi}.
\end{equation}
 $T_{c}$ approaches to the critical value of the phase transition of the background spacetime.
In the case of the bald black hole, $T_{c}$ approaches to this value for large $k$, too.
It can be also confirmed from the numerical results for $\frac{T_c}{q\mu}$ at large $\frac{k}{q\mu}$ (see Fig.~\ref{fig:tc-kcompare}).
This result illustrates the critical temperature is directly proportional to the momentum dissipation strength in the $O_1$ theory.

Next, we consider the $O_2$ theory in each background geometry.
In the bald black hole, we employ a simple trial function with one parameter,
\begin{equation}
    F_{\alpha}= z (1+ \alpha z).\quad (O_{2}\text{-theory, bald})
\end{equation}
In this case, the curve is parameterized by $k/r_{h}$ in $0<k/r_{h}<\sqrt{6}$.
In the dilatonic black hole, however, this simple trial function does not work well.
We find that the following trial function with one parameter $\alpha$, gives rational results for the wider range of $k/\mu$ in this case.
It is given by
\begin{equation}\label{imptrialf}
	F_{\alpha}(z)=
    z\left( 1+\alpha z\right)\left( 1- \frac{Pz}{r_{h}}\right) ^{-\frac{\sqrt{3}}{2}\tau}.~(O_{2}\text{-theory, dilatonic})
\end{equation}
This trial function satisfies $ F(0)=0 $ and $ F^{\prime}(0)=1 $ so it is suitable for studying the $O_2$ theory.%
\footnote{
   One can also consider $F_{\alpha} = \frac{z\left( 1+\alpha z\right)}{1 + z} \left( 1- \frac{Pz}{r_{h}}\right) ^{-\frac{\sqrt{3}}{2}\tau}$, which gives good agreement too.
}
The exponential factor $ -\frac{\sqrt{3}}{2}\tau $ is just same as those involved in the coupling term $ B(\phi) $, see Eq.~\eqref{couplingfactor}.
Figure \ref{fig:tc-kimprove} shows the results of the phase boundaries in the $\frac{T_c}{q\mu}$--$\frac{k}{q\mu}$ plane obtained by the approximation and the numerical method, for a comparison.
The approximation almost agrees with the numerical result.
Unlike the $O_1$ theory, the curves cross the line of $T=k/(2\sqrt{2}\pi)$, and $T_{c}$ reaches zero if $\tau$ is small.
In the case of the dilatonic black hole, the curves are parameterized by $P/r_h$, but it can take a value in $0<P/r_h<1$ in the $O_2$ theory.
At $P/r_h = 0$, the result does not depend on $\tau$ which is located at $(\frac{k}{q\mu},\frac{T_c}{q\mu}) \approx (0.41,0.049)$.
For large enough $\tau$, $T_{c}$ linearly depends on $k$ but the coefficient is different from Eq.~(\ref{eq:scaling_Tc-k}).
For $\tau=1/\sqrt{3}$, we numerically obtain $T_{c} \approx 0.0193 k$ for large $k$.
Such behavior in $T<k/(2\sqrt{2}\pi)$ is qualitatively same as those studied in Ref.~\cite{Homes' law}.

Finally, we show the correct phase boundaries for each theory.
Combining the results in Fig.~\ref{fig:tc-kcompare} for the $O_1$ theory, and Fig.~\ref{fig:tc-kimprove} for the $O_2$ theory, we obtain the phase boundaries as Fig.~\ref{fig:Tc-k_new}.
According to the critical value of $T = k/(2\sqrt{2}\pi)$, which is shown as the dotted line, we have switched the using background geometry and the corresponding results.
Note that the purple curves in $T > k/(2\sqrt{2}\pi)$ are independent of the choice of $\tau$ because $\tau$ represents a coupling with the dilaton.
For the $O_1$ theory, we employed only the purple curve corresponding to the bald black hole because all curves are located in $T>k/(2\sqrt{2}\pi)$.
We also provide further comparisons with the full analysis in Appendix \ref{appendixa}.
The calculation of the critical temperature $T_c$  in this study demonstrates that employing the probe limit yields coherent and meaningful results by duly considering the phase transition dynamics within the background gravity sector.

For both the $O_1$ theory and the $O_2$ theory with large $\tau$, it is revealed that $T_c$
  increases as the parameter $k$ grows at high  $k$ regimes. Specifically, for the $O_1$
  theory, the large-$k$ behavior can be effectively approximated using Eq.~(\ref{eq:scaling_Tc-k}). Although initially, this positive correlation between $T_c$
  and $k$, which signifies the strength of momentum dissipation, may seem counter-intuitive, such a trend has been consistently observed across various holographic models incorporating 
$k$ \cite{Homes' law, tomas}, as previously discussed. In light of these findings, a future discourse on the validity and implications of this scaling relationship between 
$T_c$
  and 
$k$ from the perspective of condensed matter physics would indeed be valuable and insightful. It remains to be explored whether this phenomenon aligns with physical expectations or presents novel challenges to our current understanding.

\subsection{Condensation}
In the superconducting phase, the charged scalar has nontrivial profiles below $T=T_c$.
The scalar profile is obtained by solving the nonlinear ODEs for $\Phi$ and $A_t$.
We conclude this section by showing the numerical results of the condensations for $O_1$ and the $O_2$ theory.
Here, we set $q=1$ but the choice of $q$ does not affect the results in the probe limit since it can be scaled out.

\begin{figure}[htbp]
    \centering
    \includegraphics[width=8cm]{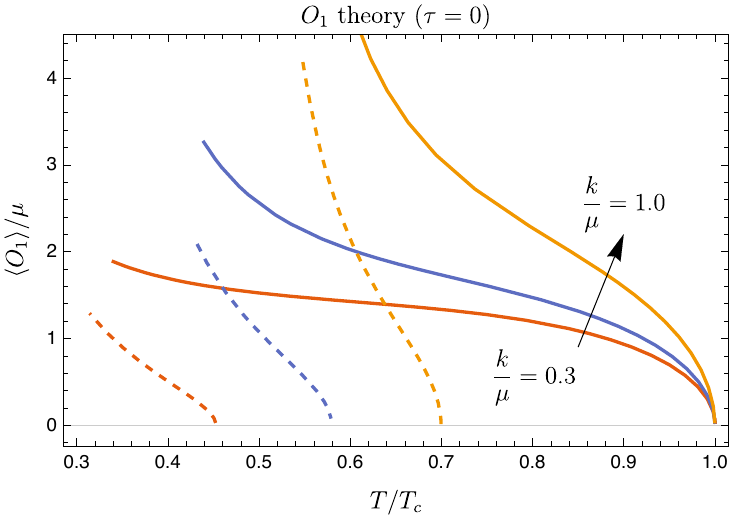}
    \includegraphics[width=8cm]{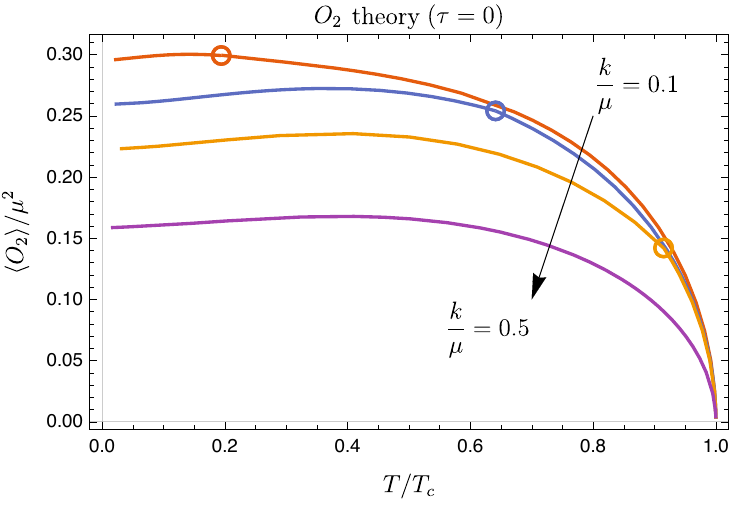}
    \caption{
        Condensation as functions of $T$ for various $k/\mu$.
        We set $\tau=0$.
        (left)
        For the $O_1$ theory.
        The solid and dashed curves show the results in the bald and the dilatonic black hole, respectively.
        $\expval{O_1}/\mu$ grows as the temperature decreases.
        We obtain the data for $T>k/(2\sqrt{2}\pi)$, i.e., the solid curves are favored.
        The solid (dashed) curves correspond to $k/\mu = 0.3, 0.6$ and $1.0$ from bottom to top.
        (right) For the $O_2$ theory.
        The curves correspond to $k/\mu = 0.1, 0.3, 0.4$ and $0.5$, from top to bottom.
        The small circles on the curves indicates the critical point at $T=k/(2\sqrt{2}\pi)$.
        Below/above the temperature of this point, the data is obtained with the dilatonic/bald black hole, respectively.
        For $k/\mu=0.5$, all data are obtained in the dilatonic black hole because $T_{c}<k/(2\sqrt{2}\pi)$ in this case.
        Unlike the $O_1$ theory, $\expval{O_2}/\mu^2$ remains finite at $T\to0$.
    }
    \label{fig:Oi-T}
\end{figure}

Figure \ref{fig:Oi-T} shows the condensates $\expval{O_i}$ as functions of $T$ for $O_i$ theories.
We set $\tau=0$ for simplicity.
To obtain this figure, we have changed the using background geometry at the phase transition point $T=k/(2\sqrt{2}\pi)$.
We show this point as a small circle on the curve, if it exists below the superconducting phase transition point.
The condensate $\expval{O_1}$ grows as temperature decreases, whereas $\expval{O_2}$ remains finite at $T\to 0$.
In particular, there is a lower bound for the possible $T$ in the $O_1$ theory.
This behavior is same as those studied in \cite{bhs}, and is a limitation in the probe limit analysis.%
\footnote{
    In the $O_2$ theory with $\tau=0$, $T\to0$ limit is possible.
}
Since the lowest temperature is always larger than $T=k/(2\sqrt{2}\pi)$ for the $O_1$ theory, the results in the bald black hole, denoted as solid curves, are favored.
We expect that the low-temperature divergence of $\expval{O_1}$ will be cured by considering beyond the probe limit, similarly to Ref.~\cite{Hartnoll:2008kx}.
One can see that $\expval{O_1}$ is enhanced by increasing $k$ for fixed $\mu$.
On the other hand, $\expval{O_2}$ is suppressed as $k$ increases.
These behaviors of the condensates are consistent with the relation between $T_c$ and $k$ for each theory, shown as Fig.~\ref{fig:Tc-k_new}.

\section{Linear dependence of \texorpdfstring{$T_{c}$}{Tc} on the superfluid density}\label{section4}
In Ref.~\cite{copperoxides}, the authors experimentally reveal the dependence of critical temperature ($T_c$) in overdoped copper oxide superconductors, specifically La$_{2-x}$Sr$_{x}$CuO$_{4}$, on their superfluid density. Using atomic layer molecular beam epitaxy, homogeneous single-crystal films with varying doping levels were synthesized to thoroughly investigate the entire overdoped region. The team measured magnetic penetration depth and zero temperature phase stiffness with high precision across thousands of samples. Results showed a consistent linear decrease of phase stiffness with temperature at all doping levels.
In this section, we compute the AC conductivity in our model to investigate the relation between $T_{c}$ and the zero-temperature superfluid density $n_{s}(0)$.
The exact AC conductivity in the dilatonic black hole, in which corresponds to the normal phase, was studied in Ref.~\cite{Ren:2021rhx}.
The superfluid density $n_{s}$ can be read from the AC conductivity in the superconducting phase.
We numerically compute the AC conductivity in the superconducting phase.

In order to compute the AC conductivity, we need to solve Eq.~\eqref{maxwelleq} for the following perturbation ansatz
\begin{equation}
    A = A_t(r) dt + e^{-i \omega t } A_{x}(r) dx,
\end{equation}
where $A_{x}(r)$ is a small perturbation field.
Linearizing Eq.~\eqref{maxwelleq} about $A_x(r)$, we obtain
\begin{equation}\label{Axeq}
	A_{x}^{\prime\prime}(r)+\left(\frac{f^{\prime}(r)}{f(r)}+\frac{\phi^{\prime}(r)}{\sqrt{3}}\right)A_{x}^{\prime}(r)+\left(\frac{\omega^{2}}{f(r)^{2}}-\frac{2q^{2}e^{-\frac{\phi(r)}{\sqrt{3}}}}{f(r)}\Phi(r)^{2}\right)A_{x}(r)=0 \ .
\end{equation}
In the $z$-coordinate, it becomes
\begin{equation}\label{Axz}	A_{x}^{\prime\prime}(z)+\left(\frac{2}{z}+\frac{f^{\prime}(z)}{f(z)}+\frac{\phi^{\prime}(z)}{\sqrt{3}}\right)A_{x}^{\prime}(z)+\frac{r_{h}^{2}\left(\omega^{2}-2e^{-\frac{\phi(z)}{\sqrt{3}}}q^{2}f(z)\Phi(z)^{2}\right)}{z^{4}f(z)^{2}}A_{x}(z)=0.
\end{equation}
To obtain the physical result of the AC conductivity, we impose the infalling-wave boundary condition at the black hole horizon.
From the equation of motion, the infalling-wave solution must have the near horizon behavior written as
\begin{equation}
    A_x(z) = \left(
        1 - z
    \right)^{\frac{-i\omega}{4\pi T}} G(z),
\end{equation}
where $G(z)$ is a regular function at the horizon $z=1$.
According to the prescription of the AdS/CFT correspondence with finite temperatures \cite{Son:2002sd}, we can compute the AC conductivity by
\begin{equation}\label{eq:conductivity}
    \sigma(\omega) =
    -
    \frac{1}{i\omega}
    \lim_{r\to\infty}
    \frac{r^2 \partial_{r} A_x(r)}{A_x(r)}
    =
    \frac{r_h}{i\omega}
    \lim_{z\to0}\frac{\partial_{z} A_x(z)}{A_x(z)}.
\end{equation}
In the following, we study the AC conductivity in our model by using both the numerical and the analytical approximate methods in each phase.

\subsection{AC conductivity in the normal phase}\label{sec:AC_cond_normal}
We now briefly review the derivation of the corresponding analytic AC conductivity in the normal phase.
The normal phase is described by $\Phi(r) = 0$, then Eq.~\eqref{Axeq} becomes
\begin{equation}
	A_{x}^{\prime\prime}(r)+\left(\frac{f^{\prime}(r)}{f(r)}+\frac{\phi^{\prime}(r)}{\sqrt{3}}\right)A_{x}^{\prime}(r)+\frac{\omega^{2}}{f(r)^{2}}A_{x}(r)=0.
\end{equation}
Using the dilatonic black hole geometry (\ref{eq:background_geometry}), the equation has regular singular points at $r=\pm r_h$ and $P$.
The solution is given by
\begin{equation}
    A_x(r) = C_{0}
    \left(
        \frac{r - r_h}{r + r_h}
    \right)^{-i\omega/4\pi T}
    {}_{2}F_{1}\left(
        \tilde{a},~ \tilde{b};~ \tilde{c};~
        \tilde{x}
        \frac{r - r_h}{r + r_h}
    \right),
\end{equation}
where $C_{0}$ is a normalization constant, and
\begin{equation}
\begin{gathered}
    \tilde{a} =
    -\frac{i\omega }{2r_{h}}\left(
        \frac{1}{\sqrt{1-P/r_h}}
        -\frac{1}{\sqrt{1+P/r_h}}
    \right),\quad
    \tilde{b} =
    -\frac{i\omega }{2 r_{h}}\left(
        \frac{1}{\sqrt{1-P/r_h}}
        +\frac{1}{\sqrt{1+P/r_h}}
    \right),\\
    \tilde{c} =
    1-\frac{i\omega/r_h}{\sqrt{1-P/r_h}},\quad
    \tilde{x} = \frac{P+r_h}{P-r_h}.
\end{gathered}
\end{equation}
Note that this solution is exactly same as those obtained in Ref.~\cite{Ren:2021rhx} for the 3-charge black hole.
The form of the expression can be exchanged by using the property of the hypergeometric function.
Using Eq.~\eqref{eq:conductivity}, the AC conductivity is obtained as
\begin{equation}\label{eq:normal_AC_cond}
    \sigma(\omega)
    =
    \frac{1}{\sqrt{1-P/r_h}} - \frac{r_h}{i \omega}
    \frac{2 \tilde{a}\tilde{b}\tilde{x}}{\tilde{c}}
    \times
    \frac{
        {}_{2}F_{1}(
            1+\tilde{a},~
            1+\tilde{b};~
            1+\tilde{c};~
            \tilde{x}
        )
        }{
        {}_{2}F_{1}(
            \tilde{a},~
            \tilde{b};~
            \tilde{c};~
            \tilde{x}
        )
    }.
\end{equation}
We show the results of Eq.~\eqref{eq:normal_AC_cond} in some cases in Fig.~\ref{fig:conductivity_near_Tc}.
From this expression, we can read off the DC conductivity as
\begin{equation}\label{sigmaDC}
    \sigma_{\text{DC}} = \lim_{\omega\to0} \sigma(\omega)
    =
    \frac{1}{\sqrt{1-P/r_h}}
    =
    \frac{1}{2\sqrt{2}\pi} \frac{k}{T}.
\end{equation}
The result agrees with the DC conductivity obtained in Refs.~\cite{Jeong:2018tua, Homes' law} with vanishing chemical potential, i.e., in the neutral limit.
To obtain the $\omega$-dependent AC conductivity and the $T$-dependent DC conductivity, the nonzero dilaton and the coupling between the dilaton and the Maxwell field are important \cite{Ren:2021rhx}.
If the theory has the S-duality, the conductivity becomes constant as those in the bald black hole that we will show next.
For more details about this point, see Ref.~\cite{Herzog:2007ij} and section 3.4.6 of Ref.~\cite{Hartnoll:2016apf}.
We can say that the presence of the non-zero dilaton plays a significant role for the emergence of the linear-$T$ resistivity here.

The above result is valid only during the dilatonic black hole (\ref{eq:background_geometry}) is the true ground state at low temperature.
If the bald black hole (\ref{eq:bald_black_hole}) becomes the ground state, the infalling-wave solution for $A_{x}$ in this geometry is obtained as
\begin{equation}
\begin{gathered}
    A_{x}(z) = C_{0} \left(
        1 - z
    \right)^{-\frac{i\omega}{4\pi T}}
    \left(
        2 + 2 z + (2 - \tilde{k}^2) z^2
    \right)^{\frac{i \omega}{8 \pi T}}
    \left(
        \frac{
            i \sqrt{3- 2 \tilde{k}^2} - 1 -(2- \tilde{k}^2)z
        }{
            i \sqrt{3- 2 \tilde{k}^2} + 1 +(2- \tilde{k}^2)z
        }
    \right)^{\lambda'},\\
    \lambda' =
    \frac{3-\tilde{k}^2}{\sqrt{3 - 2 \tilde{k}^2}}
    \frac{\omega}{8\pi T},
\end{gathered}
\end{equation}
where $\tilde{k} = k/r_{h}$.
Note that $T$ is given by Eq.~(\ref{eq:T_bald}) in the bald black hole.
In this case, however, one obtains the conductivity as $\sigma(\omega) = 1$ for any $\omega$.
Thus, the DC conductivity is also given by $\sigma_{\text{DC}} = 1$.
As a result, the (DC) resistivity behaves
\begin{equation}\label{eq:resistivity_final}
    \rho_{\text{DC}} = \frac{1}{\sigma_{\text{DC}}} = 
    \begin{cases}
        0 & T\leq T_{c}\\
        2\sqrt{2}\pi T / k & T_{c} < T \leq k/(2\sqrt{2}\pi) \\
        1 & k/(2\sqrt{2}\pi) < T
    \end{cases}.
\end{equation}
If $T_{c}$ is larger than $k/(2\sqrt{2}\pi)$, the middle regime disappears.
$\rho_{\text{DC}}$ does not jump at $T=k/(2\sqrt{2}\pi)$ because Eq.~\eqref{sigmaDC} becomes $1$ there.
\rev{
The behavior of Eq.~\eqref{eq:resistivity_final} is qualitatively same as the result in the fully-backreacted analysis \cite{Jeong:2018tua}, although $\mu$ dependence is dropped out due to the probe limit.
It implies that the probe limit analysis still can capture several significant property of this model, such as the linear-$T$ resistivity.
}

\subsection{AC conductivity in the SC phase and the superfluid density}
\begin{figure}[htbp]
    \centering
    \includegraphics[width=16cm]{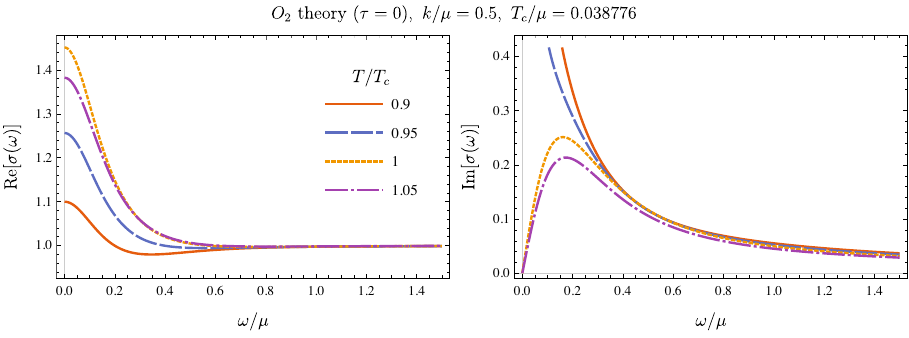}
    \caption{
        AC conductivity for $k/\mu=0.5$ around the critical temperature in the $O_2$ theory.
        Above $T_c$, the result is given by Eq.~\eqref{eq:normal_AC_cond}.
        Below $T_c$, the result is obtained numerically.
        One can see the $1/\omega$ dependence implying the superconductivity in the imaginary part of the conductivity for $T<T_{c}$.
    }
    \label{fig:conductivity_near_Tc}
\end{figure}

In the superconducting phase, we need the numerical analysis to study the AC conductivity basically.
With the benefit of the simplification in the probe limit, we can easily compute it by using the standard procedure in the holography.
Figure \ref{fig:conductivity_near_Tc} shows the AC conductivity in the $O_2$ theory with $\tau=0$ for $k/\mu=0.5$ and various $T$ near $T_{c}$.
Note that $T_{c}<k/(2\sqrt{2}\pi)$ in this case.
We can use the dilatonic black hole as the background geometry.
In the superconducting phase, below $T_c$, we can observe $1/\omega$ dependence in the imaginary part of the AC conductivity, which implies the delta peak in the real part.

\begin{figure}
    \centering
    \includegraphics[width=8cm]{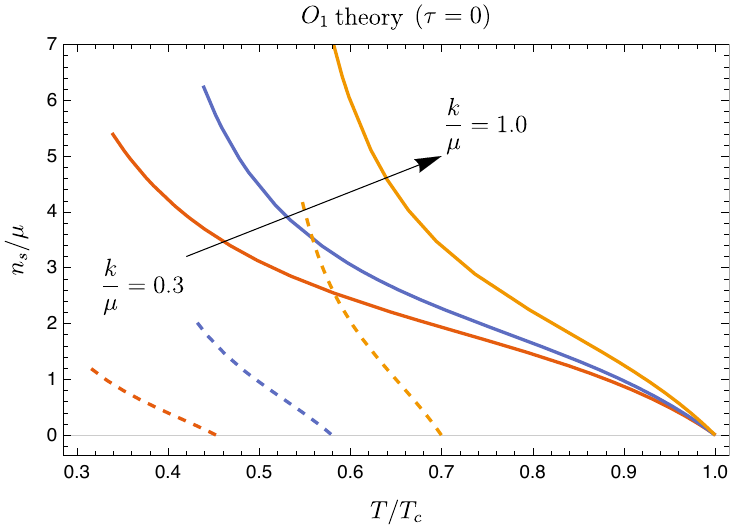}
    \includegraphics[width=8cm]{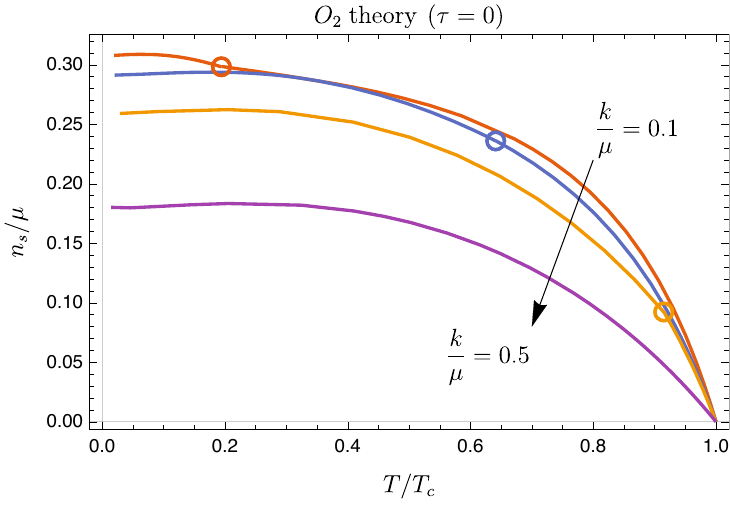}
    \caption{
        Superfluid density as functions of $T$ for various $k/\mu$.
        We set $\tau=0$.
        (left) For the $O_1$ theory.
        (right) For the $O_2$ theory.
        The curves corresponds to those shown in Fig.~\ref{fig:Oi-T}.
        The behavior of $n_{s}$ is very similar to the behavior of $\expval{O_i}$.
        A different point is that $n_{s}(T)$ goes zero linearly near $T=T_{c}$.
    }
    \label{fig:ns-T}
\end{figure}

\rev{
Another interesting observable in the superconducting phase is a superfluid density $n_s$.
It can be read from the low-frequency expansion of the AC conductivity as
\begin{equation}\label{eq:def_ns}
    \sigma(\omega) = \pi n_s \delta(\omega) + \frac{i n_s}{\omega} + \order{\omega}.
\end{equation}
More precisely, $n_{s}$ is proportional to the superfluid density.}%
\footnote{
    $n_{s}$ defined in (\ref{eq:def_ns}) may be called as the phase stiffness rather than the superfluid density in some studies.
    In fact, the dimension of $n_{s}$ is $1$, which is different from the dimension of the number density in $(2+1)$d system.
    The number density of the super-fluid components, $\rho_{s}$, might be given by $n_{s} = \rho_{s}/m^{*}$, where $m^{*}$ denotes the effective mass of the charged carrier.
    In the holographic superconductor models, one may take $m^{*}=\mu$.
    See, e.g., section 6.4.1 of Ref.~\cite{Hartnoll:2016apf} for more details.
}
Remark that the real and the imaginary parts of the conductivity are related with each other by the demand of the causality.
Figure \ref{fig:ns-T} shows the numerical results of $n_s$ as functions of $T$ for various $k$.
The behavior of $n_s$ is similar to $\langle O_i\rangle$ at low temperatures, see Fig.~\ref{fig:Oi-T}.
Unlike Fig.~\ref{fig:Oi-T}, we can see from the numerical results that $n_{s}(T)$ has linear behavior near $T=T_{c}$.

\begin{figure}
    \centering
    \includegraphics[width=10cm]{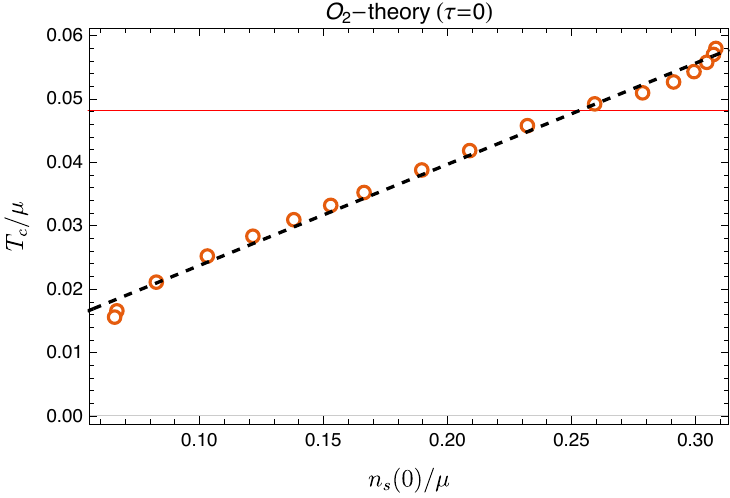}
    \caption{Relation between $T_c$ and $n_s(0)$ in the $O_2$ theory with $\tau=0$.
    $n_{s}(0)$ is defined by $n_{s}(0)=\lim_{T\to0}n_{s}(T)$.
    We find $n_{s}(0)$ and $T_c$ exhibit the linear relation roughly.
    The black dashed line shows the fit \eqref{eq:ns-Tc_scaling}.
    The horizontal red line shows the critical point $T_{c}/\mu = 0.048218$ where $2\pi T_{c}= k/\sqrt{2}$ is satisfied.
    Above this line, we use $T_{c}$ obtained by using the bald black hole, which is thermodynamically favored.
    }
    \label{ns-Tc}
\end{figure}

In the $O_2$ theory with $\tau = 0$, $n_s$ remains finite in the vicinity of $T=0$ even in the probe limit analysis.
Figure \ref{ns-Tc} shows the relation between $n_s$ at $T=0$ and $T_{c}$ in the $O_2$ theory with $\tau=0$.
We find $n_s(0)$ and $T_c$ have weakly linear relation as shown in this figure.
\rev{
The relation can be fitted by
\begin{equation}\label{eq:ns-Tc_scaling}
    T_{c} \sim 0.159\, n_{s}(0) + 0.00778\,\mu,
\end{equation}
which is shown as the dashed line in Fig.~\ref{ns-Tc}.
}
\rev{
A similar relation has been also observed in the overdoped side of a copper-oxide film \cite{copperoxides}.
}
On the other hand, there is a universal scaling law between $n_{s}(0)$ and $T_{c}$ known as Homes' law given by $n_{s}(0) \propto \sigma_{\text{DC}}(T_{c})\times T_{c}$ in unconventional superconductors \cite{Homes:2004,Homes:2005}.
Our result shows the linear relation between $n_{s}(0)$ and $T_{c}$ but $\sigma_{\text{DC}}$ given by Eq.~\eqref{sigmaDC} must enter here.
Hence, the Homes' law is not held.
This is stirring result that may imply the $O_2$ theory with $\tau=0$ of our model can describe LSCO thin films.

We have to remark some points on Fig.~\ref{ns-Tc}.
First, it is crucial to note that the precise zero-temperature limit in our geometry presents a subtle issue because the locations of the horizon and the singularity converge at a single point.
To estimate $n_s(0)$, we rely on numerical AC conductivity data for a near-zero temperature scenario with $P/r_h = 0.99999$ because $P/r_h=1$ corresponds to $T=0$ from Eq.~\eqref{eq:temperature}. The empirical values are obtained from solutions where the temperature is approximately an order of $T/T_{c} \sim 10^{-3}$.

Secondly, for very small critical temperatures  $T_{c}$,  the relationship between $T_{c}(k)$ and $k$  becomes multivalued within a narrow range of $k$. Consequently, there is ambiguity about which value of  $\left.n_{s}(0)\right|_{k}$  corresponds to which specific $T_{c}(k)$.  Due to this complexity, we have omitted the small $T_c$ data from Fig.~\ref{ns-Tc}.

Lastly, the background spacetime undergoes a phase transition. In Fig.~\ref{ns-Tc}, the red horizontal line denotes the critical threshold corresponding to $T=k/(2\sqrt{2}\pi)$. Above this line, the bald black hole described by Eq.~(\ref{eq:bald_black_hole}) represents the stable ground state. The data points presented in Fig.~\ref{ns-Tc} inherently account for this phase transition.

\section{Discussion and Conclusion}\label{section5}
To unravel the intricate scaling relations in holographic superconductors, we investigate the relation among the momentum dissipation, the critical temperature, and the zero-temperature phase stiffness in the Gubser-Rocha model within the probe limit framework.
This approach has enabled us to examine various fundamental aspects of the holographic superconductor model.
We have shown the several properties of the holographic superconductor model, e.g., the critical temperature $T_{c}$, the condensations $\expval{O_i}$, and also the AC conductivity $\sigma(\omega)$ in both phases.
In the analysis of the critical temperature, we have found the scaling relation \eqref{eq:scaling_Tc-k} between $T_c$ and $k$ for the $O_1$ theory.
We have also studied the relation between $T_{c}$ and the superfluid density $n_{s}$ at zero temperature in the $O_2$ theory as shown in Fig.~\ref{ns-Tc}.
We found the roughly linear relation, which can be fitted well as Eq.~(\ref{eq:ns-Tc_scaling}) for $\tau=0$.
Similar relations have been observed in real experiments, as we will discuss below.

\rev{
A similar linear relation between $T_{c}$ and $n_{s}(0)$ has also been observed in a recent experiment in a cooper-oxide film \cite{copperoxides}.
Our result may imply that the $O_2$ theory with $\tau=0$ of our model can describe such cuprates.
Note that it differs from the direct proportionality in underdoped materials, known as Uemura relation \cite{Uemura:1989,Uemura:1991}.
}
On the other hand, the Homes' law \cite{Homes:2004,Homes:2005} is known as the universal scaling relation, which works regardless of whether overdoped or underdoped materials.
This relation is expressed as $n_{s}(0) \propto \sigma_{\text{DC}}(T_{c}) T_{c}$.
The Homes' law in the holographic superconductor model was tested in Ref.~\cite{Homes' law} in the analysis with backreactions.
They concluded that the Homes' law holds in the $O_2$ theory with sufficiently large $\tau$ in this model.
\rev{
In our computation of the superfluid density, we focus on $\tau=0$, where they claim the Homes' law does not hold in this model.
We also remark that the material in \cite{copperoxides} also does not obey the Homes' law.
It was discussed in \cite{dordevic}.
If we attempt to check the Homes' law in this model under the probe limit, two problems arise.
First, the Homes' law involves the normal phase DC conductivity at $T=T_c$, which is inaccurate in the probe limit.
Second, the superfluid density does not converge at low temperatures for $\tau>\tau_{c}$, like in the $O_1$ theory, so we cannot obtain $n_{s}(0)$.
}
Even though the Homes' law does not cover all superconductors, it still valid for various conventional and unconventional superconductors.
On the other hand, only the specific holographic model admits the Homes' law in the limited situations.
This mystery of the discrepancy still need to be understood in the future studies.

Our primary emphasis lies on the superfluid density; however, the behavior of the normal fluid density plays a vital role in high-temperature superconductors.
Recent studies \cite{Gouteraux:2019kuy,Gouteraux:2020asq} have uncovered that in certain classes of holographic superconductors, the normal fluid density persists even down to absolute zero temperature --- a phenomenon starkly contrasting the predictions of BCS theory, wherein the normal fluid density vanishes at zero temperature.
According to \cite{Gouteraux:2020asq}, this unconventional persistence occurs when the dynamical exponent $z$ exceeds $d+2$, where $d+1$ represents the spacetime dimension of the boundary theory.
Given that the Gubser-Rocha black hole corresponds to $z \rightarrow \infty$, our model likely falls within this category. 

Within our setup, the strength of momentum dissipation (or disorder) represented by $k$ introduces an additional parameter to the system compared to the standard holographic superconductor model (as reviewed in \cite{bm}). While the role of 
$k$ is clear in breaking the translational symmetry explicitly in a full analysis with backreactions, its direct experimental correspondence remains uncertain.
Typically, translational symmetry is indeed broken in real-world materials, but the intensity of this breakage can vary with changes in other parameters.  In experiments involving cuprate superconductors, for instance, hole doping $p$ significantly impacts the material properties.
Cuprates exhibit superconductivity within a specific doping range, leading to a characteristic dome-shaped phase diagram along the $p$-axis.
It is conjectured that changes in doping levels also lead to variations in 
$k$.
Another factor that can induce momentum dissipation is the strength of defects, which can be manipulated through processes like irradiation.
It's worth noting that translational invariance can also be broken by considering a spatially varying chemical potential $\mu(x)$
 \cite{Horowitz:2012ky}.
Works such as \cite{Arean:2013mta, Arean:2015sqa} have introduced charged disorder via a random profile of $\mu(x)$ within the probe limit.
The presence of $ k$ can be viewed as an analogy to these mechanisms, albeit independent of charge.
Establishing a precise connection between $ k$ and experimentally controllable parameters would be both intriguing and significant.

\rev{
It is important to acknowledge the limitations imposed by adopting the probe limit in our analysis.
Firstly, even when setting $k=0$, the conductivity in the normal phase does not vanish, contrary to what one might expect.
This anomaly arises due to the probe limit itself, which inherently leads to a broken translational symmetry.
The explicit dependence on the charge density is also dropped out from the conductivity.
Secondly, the condensation and the superfluid density sometimes do not converge at low temperatures.
Indeed, we encountered a challenge in obtaining solutions for the superconducting phase below the temperature threshold of $T=k/(2\sqrt{2}\pi)$ within the $O_1$ theory.
It also happens in the $O_{2}$ theory, when we set large $\tau$.
}
In Appendix \ref{appendixa}, we have numerically verified that the phase boundary calculated under the probe limit does indeed align with the full result when considering a large value of $q$ along with appropriate rescaling.
While the probe limit method has the above limitations, it simultaneously simplifies the model's analysis considerably, making it a practical starting point.
Future work, encompassing a more comprehensive treatment of backreactions, is necessary to overcome these constraints and achieve a more accurate depiction of the system's behavior, especially at low temperatures and for the $O_1$ theory.

\rev{
In this paper, we provided the first comprehensive analysis of the probe limit in this model, as far as we know.
Note that the AC conductivity calculation with the black hole solution \eqref{eq:background_geometry} has been examined in \cite{Ren:2021rhx}, and the same neutral geometry with $k=0$ was briefly mentioned in Appendix B of \cite{Ren:2022qkr}.
}
Expanding upon this groundwork, it would be intriguing to tackle more complex scenarios that are challenging to study in a full backreaction analysis, such as spatially inhomogeneous solutions \cite{Keranen:2009vi} or the temporal evolution of the system \cite{Zeng:2016gqj,Yang:2023dvk}.
The inclusion of a running dilaton can yield distinct outcomes from standard holographic superconductor setups in the SAdS$_4$ spacetime, although it is essential to remember that the dilatonic black hole becomes the ground state only when the temperature satisfies $T<k/(2\sqrt{2}\pi)$.

Moreover, exploring the impact of magnetic fields in our model promises to be equally fascinating. A dyonic black hole geometry in the Gubser-Rocha model serves as the normal phase vacuum, as recently investigated in Ref.~\cite{Ge:2023yom}. However, in the probe limit, our neutral black hole geometry remains applicable since the Maxwell fields are uncoupled from gravity, as evidenced in analogous studies within the SAdS$_4$
  context \cite{Ge:2010aa, Domenech:2010nf}. Here, the effect of an external magnetic field solely permeates the system through the coupled Maxwell-scalar sector under the probe limit conditions. By introducing such a magnetic field, it is anticipated that the phase boundaries and condensates will be altered accordingly. Of particular significance is the role played by the dynamical $U(1)$ gauge field in capturing various magnetic field effects \cite{Domenech:2010nf, Salvio:2012at, Salvio:2013jia}. Holographic superconductors featuring dynamical $U(1)$ gauge fields are known to emulate type II superconductors, and the penetration depth computed in their presence constitutes a crucial quantity that can be more directly compared to experimental findings.
  These areas remain ripe for future research endeavors.

\section*{Acknowledgments}
We would like to thank Sang-Jin Sin, Matteo Baggioli, Xinmao Yin,  Hui Xing, and Di-fan Zhou for their helpful comments.
We are grateful to Hyun-Sik Jeong for suggestive comments, especially about the thermodynamic stability of the background black hole geometry.
This work is partly supported by NSFC, China ( Grant No. 12275166 and No. 12311540141).

\appendix
\section{Checking the correctness of the probe limit}\label{appendixa}
In this section, we compare the $T_c$--$k$ curves obtained by the probe limit analysis with those obtained for general $q$.
Firstly, we briefly review the charged black hole solutions of our model (\ref{eq:full_model}), and its thermodynamic stability.
From the full action (\ref{eq:full_model}), the equations of motion are obtained as
\begin{equation}
	\grad_{\mu}{\left(e^{\frac{\phi}{\sqrt{3}}}F^{\mu\nu}\right)}-iq\Phi^{*}\left(\partial^{\nu}-iqA^{\nu}\right)\Phi+iq\Phi\left(\partial^{\nu}+iqA^{\nu}\right)\Phi^{*}=0,
\end{equation}
\begin{equation}
	\grad^{2}{\phi}-\frac{1}{4\sqrt{3}}e^{\frac{\phi}{\sqrt{3}}}F^{2}+2\sqrt{3} \sinh\left(\frac{\phi}{\sqrt{3}}\right)-B^{\prime}(\phi)\abs{\Phi}^{2}=0,
\end{equation}
\begin{equation}
	D_{\mu}D^{\mu}\Phi-B(\phi)\Phi=0,\quad
    \grad^{2}{\psi_{I}}=0.
\end{equation}
The Einstein's equation is
\begin{align}
	R_{\mu\nu}-&\frac{1}{2}g_{\mu\nu}\left[R-\frac{1}{4}e^{\frac{\phi}{\sqrt{3}}}F^{2}-\frac{1}{2}(\partial \phi)^{2}+6 \cosh\left(\frac{\phi}{\sqrt{3}}\right)-\frac{1}{2}\sum_{I=1}^{2}\left(\partial \psi_{I}\right)^{2}-\abs{D\Phi}^{2}-B(\phi)\abs{\Phi}^{2}\right]\nonumber\\ 
	&=\frac{1}{2}e^{\frac{\phi}{\sqrt{3}}}F_{\mu\delta}F_{\nu}^{\ \delta}+\frac{1}{2}\partial_{\mu}\phi\partial_{\nu}\phi+\frac{1}{2}\sum_{I=1}^{2}\left(\partial_{\mu}\psi_{I}\partial_{\nu}\psi_{I}\right)+\frac{1}{2}\left(D_{\mu}\Phi D_{\nu}^{*}\Phi^{*}+D_{\nu}\Phi D_{\mu}^{*}\Phi^{*}\right).
\end{align}
The normal phase solution is obtained as the following charged dilatonic black hole solution \cite{Jeong:2018tua}:
\begin{equation}
	ds^2 = - f(r)dt + \frac{dr^2}{f(r)}
	+ h(r) (dx^2 + dy^2),
\end{equation}
with
\begin{subequations}
\begin{gather}
	h(r) = r^2 \left(1-\frac{P}{r}\right)^{1/2},\quad
	f(r) = h(r) \left[
		1 - \frac{k^2}{2 r^2}
		- \frac{r_h^3}{r^3}\left(
			1 - \frac{k^2}{2 r_h^2}
		\right)
	\right],\\
	\phi(r) = - \frac{\sqrt{3}}{2} \ln\left(
		1 - \frac{P}{r}
	\right),\quad \Phi = 0, \quad \psi_I = k x^I,\\
	A =
	\sqrt{3 P r_h \left(
		1 - \frac{k^2}{2 r_h^2}
	\right)}\left(1-\frac{r_h}{r}\right) dt,
\end{gather}
\end{subequations}
where $r_h$ is the location of the black hole horizon and $P$ is a physical parameter.
Unlike Eq.~(\ref{eq:background_geometry}), $r_h$ and $k$ are independent parameters here.
Note that this is the same solution as those in Ref.~\cite{Homes' law}, but the coordinate is different.
Writing the radial coordinate in \cite{Homes' law} as $\tilde{r}$, we obtain the relation between ours and their conventions as $\tilde{r}+Q = r$, $\tilde{r}_h+Q = r_h$ and $Q = P$.
The Hawking temperature and the chemical potential are related to the parameters by
\begin{equation}\label{eq:T-mu_full}
	T =
	\frac{1}{8\pi} \frac{6 r_h^2 - k^2}{r_h}
	\sqrt{1 - \frac{P}{r_h}},\quad
	\mu =
	\sqrt{3 P r_h \left(1-\frac{k^2}{2r_h^2}\right)}.
\end{equation}
One can see that the neutral limit of Eq.~(\ref{eq:background_geometry}) is achieved by setting $r_h = k/\sqrt{2}$, whereas $P=0$ leads Eq.~(\ref{eq:bald_black_hole}).
$P/r_{h} \to 1$ corresponds to the extremal limit.
For the thermodynamic analysis, the grand potential (density) of this solution is given by \cite{Caldarelli:2016nni,Kim:2017dgz}
\begin{equation}\label{eq:grandpotential_P}
    \Omega(\mu, T; k) = - r_{h}^3 \left(
        1 + \frac{1-P/r_{h}}{2} \frac{k^2}{r_{h}^2}
    \right).
\end{equation}
The grand potential implicitly depends on $\mu$ and $T$ via $P$ and $r_{h}$.

On the other hand, the model also admit the following solution: \cite{Andrade:2013gsa}
\begin{subequations}
\begin{gather}
    h(r) = r^2,\quad
    f = r^2 \left[
        1 - \frac{k^2}{2r^2}
        + \frac{r_{h}^2 \mu^2}{4 r^4}
        - \frac{r_h^3}{r^3}\left(
            1 - \frac{k^2}{2r_{h}^2} + \frac{\mu^2}{4 r_{h}^2}
        \right)
    \right],\\
    A = \mu
    \left(1 - \frac{r_{h}}{r}\right) dt,\quad
    \phi = \Phi = 0,\quad
    \psi_{I} = k x^{I},
\end{gather}
\end{subequations}
where $\mu, k$ and $r_{h}$ are integration constants, which parameterize the family of the solution.
$\mu$ is directly read as the chemical potential.
The temperature is given by
\begin{equation}\label{eq:T-mu_full_bald}
    T = \frac{r_{h}}{4\pi}\left(
       3 - \frac{k^2}{2 r_{h}^2} - \frac{\mu^2}{4 r_{h}^2}
    \right).
\end{equation}
One can see $\mu=0$ leads the neutral solution of Eq.~(\ref{eq:bald_black_hole}).
The grand potential of this solution is given by \cite{Andrade:2013gsa}
\begin{equation}\label{eq:grandpotential_bald}
    \Omega(\mu, T; k) = -r_{h}^3
    \left(
        1 + \frac{k^2}{2 r_{h}^2} + \frac{\mu^2}{4 r_{h}^2}
    \right).
\end{equation}
The grand potential implicitly depends on $T$, again.

\begin{figure}[htbp]
    \centering
    \includegraphics[width=9cm]{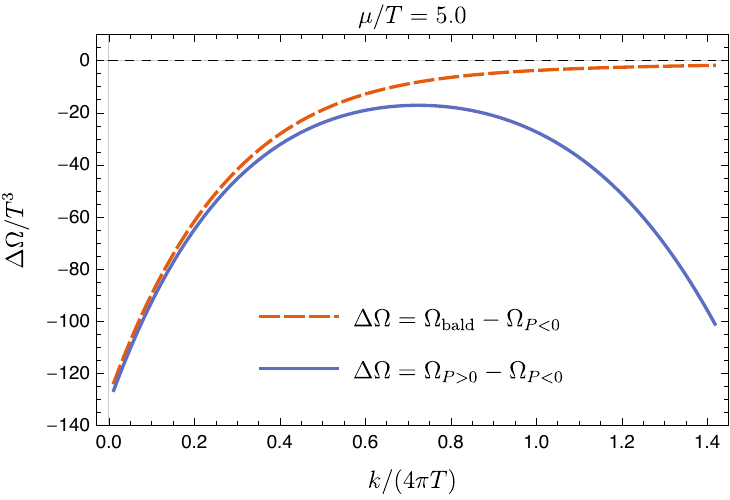}
    \caption{
        Difference of the grand potentials among the charged black hole solutions.
        We fix $\mu/T=5.0$.
        The result implies $\Omega_{P>0} < \Omega_{\text{bald}} < \Omega_{P<0}$.
    }
    \label{fig:grandpotential}
\end{figure}

Let us now consider which solution and parameter range correspond to the ground state of the normal phase.
From Eq.~(\ref{eq:T-mu_full}), there are two distinct parameter regions yielding $T,\mu >0$ for the dilatonic black hole; i) $0<P<r_{h}$ and $0< k< \sqrt{2} r_{h}$ ii) $P < 0$ and $\sqrt{2} r_{h} < k < \sqrt{6} r_{h}$.%
\footnote{
    In terms of the original parameter $\tilde{r}_{h}$, the ranges of such parameter regions become more complicated.
}
The both choice of the patches cover the entire region of the physical parameter-space like $(k/T,\mu/T)$.
On the other hand, $k<\sqrt{6 r_{h}^2 - \mu^2/2}$ is obtained from Eq.~(\ref{eq:T-mu_full_bald}) for the bald black hole.
To determine the ground state, we need to compare the grand potential among these cases.
It was investigated in \cite{Kim:2017dgz}, and the answer is that the $P>0$ patch of the dilatonic black hole is always ground state.
Figure \ref{fig:grandpotential} shows the competition of the grand potential among these solutions for fixed $\mu/T = 5.0$.
$\Omega_{\text{bald}}$ denotes Eq.~(\ref{eq:grandpotential_bald}).
$\Omega_{P>0}$ and $\Omega_{P<0}$ denote Eq.~(\ref{eq:grandpotential_P}) in the $P>0$ and $P<0$ patches, respectively.
The curves exhibit that $\Omega_{\text{bald}} - \Omega_{P<0} <0$, $\Omega_{P>0} - \Omega_{P<0} < 0$ and $\Omega_{P>0} - \Omega_{P<0} <\Omega_{\text{bald}} - \Omega_{P<0}$.
It reads $\Omega_{P>0} < \Omega_{\text{bald}} < \Omega_{P<0}$.
Therefore, we conclude that the $P>0$ patch is the ground state, which is thermodynamically favored \cite{Kim:2017dgz}.
We have checked this behavior of the grand potentials does not change for another choice of $\mu/T$.

Now, we consider the linear perturbation of the charged scalar around the normal phase solution.
The form of the equation of motion for the charged scalar is unchanged from \eqref{Phieqr} in the probe limit analysis.
In the backreacted case, the equation can no longer be written in the form of the SL problem.
We numerically solve the equation to find $T_c$ (or $\mu_c$).
In general, we can obtain multiple eigenvalues and eigenfunctions.
The solution with no node is considered as the most relevant solution to the instability.

\begin{figure}[htbp]
	\centering
	\includegraphics[width=12cm]{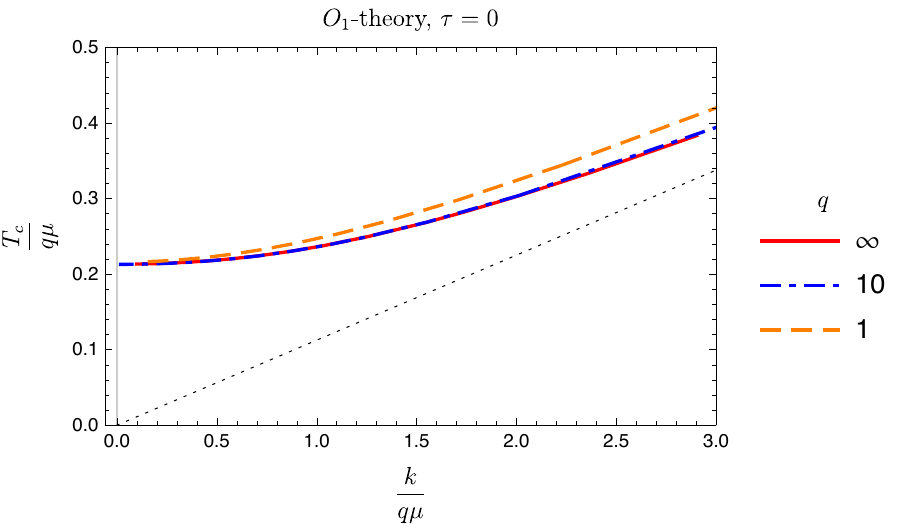}
	\caption{
    Superconducting phase boundaries of the $O_1$ theory in the $\frac{T_c}{q\mu}$--$\frac{k}{q\mu}$ plane for various $q$.
    The results for $q=\infty$ denote the results in the probe limit.
    The $q=10$ curves almost overlap with the $q=\infty$ curves.
    The dotted line shows $T_{c}=k/(2\sqrt{2}\pi)$.
    }
	\label{fig:Tc-k_O1_full}
\end{figure}

\begin{figure}[htbp]
	\centering
    \includegraphics[width=12cm]{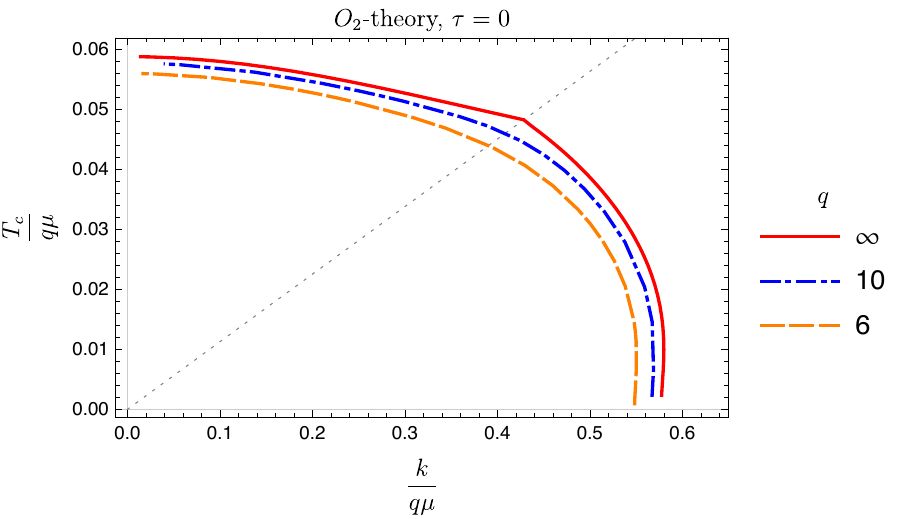}
	\caption{
    Superconducting phase boundaries of the $O_2$ theory in the $\frac{T_c}{q\mu}$--$\frac{k}{q\mu}$ plane for various $q$.
    The results for $q=\infty$ denote the results in the probe limit.
	The result for $q=6$ corresponds to one studied in Ref.~\cite{Homes' law}.
    The dotted line shows $T_{c}=k/(2\sqrt{2}\pi)$.
    }
	\label{fig:Tc-k_O2_full}
\end{figure}

Figures \ref{fig:Tc-k_O1_full} and \ref{fig:Tc-k_O2_full} show the phase boundaries in the $\frac{T}{q\mu}$--$\frac{k}{q\mu}$ plane for various $q$ in each theory.
The results for $q=\infty$ denote the results obtained by using the neutral solution \eqref{eq:background_geometry} and \eqref{eq:bald_black_hole}, i.e., the results in the probe limit.
Here, we set $\tau=0$ for simplicity, and show only the relevant results in the ground states.
In both cases, we can see that the full results approach those in the probe limit as we increase $q$.
These direct comparisons imply that we need considering the phase transition of the black hole in the probe limit.


\section{Approximate formulas of the AC conductivity in the SC phase}\label{appendix:low-T}
In this section, we derive approximate expressions of the AC conductivity within the superconducting phase where the dilatonic black hole geometry (\ref{eq:background_geometry}) is favored.
We show two results obtained by using two distinct methods developed in \cite{Herzog:2009xv, Siopsis:2010uq, Gary}.

\subsection{A method using the tortoise coordinate}
\begin{figure}
    \centering
    \includegraphics[width=10cm]{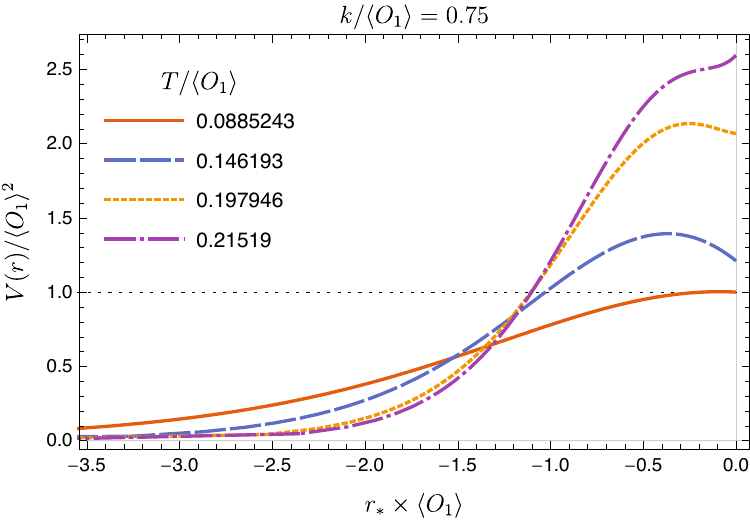}
    \caption{%
       Potential (\ref{V}) for various $T/\expval{O_2}$ in the $O_1$ theory.
       We set $q=1$ and $k/\expval{O_1}=0.75$.
       The dotted horizontal line shows $V(r)/\expval{O_1}^2=1$.
    }
    \label{potential}
\end{figure}

First, we employ the method developed in Ref.~\cite{Herzog:2009xv, Siopsis:2010uq, Gary}.
We focus on the $O_{1}$-theory.
In this case, $\Phi$ is written as
\begin{equation}
	\Phi(z)=\frac{\langle O_1 \rangle}{\sqrt{2}}\frac{z}{r_{h}}F(z),
\end{equation}
satisfying $F(0)=1$ and $F'(0)=0$.
In the superconducting phase, the vector perturbation equation is given by Eq.~\eqref{Axeq}.
Rewriting
$ A_{x}(r)=e^{-\frac{\phi(r)}{2\sqrt{3}}}B_{x}(r)$,
we obtain
\begin{equation}\label{Bx}
\begin{aligned}
	&B_{x}^{\prime\prime}(r)+\frac{f^{\prime}(r)}{f(r)}B_{x}^{\prime}(r)+\frac{\omega^{2}}{f(r)^{2}}B_{x}(r)\\
	&=
	\left[\frac{2q^{2}e^{-\frac{\phi(r)}{\sqrt{3}}}}{f(r)}\Phi(r)^{2}+\frac{1}{2\sqrt{3}}\left(\frac{f^{\prime}(r)}{f(r)}\phi^{\prime}(r)+\frac{\phi^{\prime}(r)^{2}}{2\sqrt{3}}+\phi^{\prime\prime}(r)\right)\right]B_{x}(r).
\end{aligned}
\end{equation}
Now, we consider the tortoise coordinate
\begin{equation}
	r_{*}=\int_{r_h} \frac{dr}{f(r)}=\frac{1}{2r_{h}\sqrt{1-P/r_{h}}}\text{ln}\left|r-r_{h}\right|-\frac{\sqrt{1-P/r_{h}}}{4(P-r_{h})^{2}}(r-r_{h})+\cdots.
\end{equation}
In this coordinate, the horizon is located at $r_{*}=-\infty$, and the boundary is located at $r_{*}=0$.
Note that $r_{*}$ can be written as $r_{*} \approx -z/r_{h}$ near the boundary.
By using this coordinate, Eq.~\eqref{Bx} can be written as
\begin{equation}\label{Bxtor}
	\ddot{B}_{x}(r_{*})+\omega^{2}B_{x}(r_{*})=V(r)B_{x}(r_{*}),
\end{equation}
where the dot denotes $\partial_{r_{*}}$, and $V$ is given by
\begin{equation}\label{V}
	V(r)=2q^{2}e^{-\phi/\sqrt{3}}f\Phi^{2}+\frac{f^{2}}{2\sqrt{3}}\left(\frac{f^{\prime}}{f}\phi^{\prime}+\frac{{\phi^{\prime}}^{2}}{2\sqrt{3}}+\phi^{\prime\prime}\right).
\end{equation}
The values of the potential at the horizon and boundary are
\begin{equation}
	V(r_{h})=0, \qquad V(\infty)\equiv V_{\infty}= q\expval{O_1}^{2}+\frac{3}{16}P^{2},
\end{equation}
respectively.
Figure \ref{potential} shows the shapes of the potential for various $T/\expval{O_1}$ in the $O_1$ theory.

To solve \eqref{Bxtor}, we utilize the approximate method developed in \cite{Siopsis:2010uq}.
Now, we consider replacing $V(r)$ in Eq.~\eqref{Bxtor} with a `mean value' denoted by $\bar{V}$ which is a $r$-constant.
Assuming $\omega^2 > \bar{V}$, we obtain the infalling wave solution as
\begin{equation}
	B_{x}(r_{*})=C\exp( i\sqrt{\omega^{2}-\bar{V} }r_{*} ),
\end{equation}
where $C$ is a normalization constant.
The `mean value' is defined by
\begin{equation}\label{averagev}
	 \bar{V}=\frac{\int_{-\infty}^{0}dr_{*}V(r)\left|B_{x}(r_{*})\right|^{2}}{\int_{-\infty}^{0}dr_{*}\left|B_{x}(r_{*})\right|^{2}}.
\end{equation}
The integrals diverge due to the infinite volume of the integral region.
It will be regularized by introducing a large cutoff.%
\footnote{
    In \cite{Siopsis:2010uq}, they considered it can be regularized by inserting a small imaginary part of $\omega$.
}
In this study, however, we just consider only the leading contribution at $r=\infty$:
\begin{equation}
    \bar{V} = V_{\infty} = q \expval{O_1}^2 + \frac{3}{16} P^2.
\end{equation}
This choice will be valid in a case where $r_h \ll \expval{O_1}$.
Using this result, $A_x$ is approximated as
\begin{equation}
	A_{x}(r_{*})\approx C\text{exp}\left(i\sqrt{\omega^{2}-q\langle O_{1}\rangle^{2}-\frac{3}{16}P^{2}}\ r_{*}-\frac{\phi(r)}{2\sqrt{3}}\right).
\end{equation}
According to the standard prescription of the AdS/CFT correspondence \cite{Son:2002sd}, and the Kubo formula, the conductivity is obtained as 
\begin{equation}\label{sigmaAC1}
	\sigma(\omega)\approx i\frac{P}{4\omega}+\sqrt{1-\frac{\frac{3}{16}P^{2}+q^{2}\langle O_{1}\rangle^{2}}{\omega^{2}}}.
\end{equation}
If $P=0$, the above result reduces to the approximate formula in the original model of the holographic superconductor, which is given by \cite{Gary,Siopsis:2010uq}
\begin{equation}\label{eq:AC_cond_simple}
    \sigma(\omega) \approx \sqrt{1- \frac{q^2 \expval{O_1}^2}{\omega^2}}.
\end{equation}
Although the infalling boundary condition is only valid for $\abs{\omega} > \sqrt{\bar{V}}$, the approximate formula can fit even in a small $\omega$ region.
Expanding Eq.~\eqref{sigmaAC1} in small $\omega$, we obtain the superfluid density as
\begin{equation}\label{ns1}
	n_{s}\approx\frac{P}{4}+\sqrt{\frac{3}{16}P^{2}+q^2\langle O_{1}\rangle^{2}} , \quad T\ll \langle O_1\rangle.
\end{equation}
Note that $P$ is written in terms of $k$ and $T$ as $P = k/\sqrt{2} - 4\sqrt{2}\pi^2 T^2/k$.
Thus, this is a function of $T, k$ and $\expval{O_1}$.

\begin{figure}[htbp]
    \centering
    \includegraphics[width=16cm]{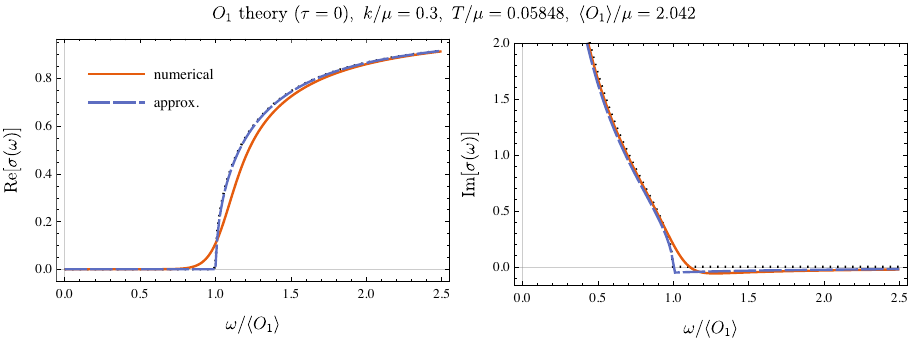}
    \caption{
        AC conductivity for $k/\mu=0.3$ of the $O_1$ theory in the dilatonic black hole.
        The red curve shows the numerical result, and the blue dashed curve shows the approximation given by Eq.~\eqref{sigmaAC1}.
        The black dotted curve, which is almost overlapped by the blue dashed curve, shows Eq.~\eqref{eq:AC_cond_simple}.
    }
    \label{fig:conductivity_low_T}
\end{figure}

For a comparison, we show Eqs.~\eqref{sigmaAC1} and \eqref{eq:AC_cond_simple} with the numerical result, for a specific background solution, in Fig.~\ref{fig:conductivity_low_T}.
The background solution is parameterized by $P/r_h = -2$, and its temperature is almost $T/T_{c} = 0.60$.
While the temperature is not sufficiently low, the AC conductivity roughly exhibits the gap.
In this case, the gap of the value is almost given by $\omega \approx \expval{O_1}$, and the curves for Eq.~\eqref{sigmaAC1} and Eq.~\eqref{eq:AC_cond_simple} are almost overlapping.
Eq.~\eqref{ns1} also gives $n_{s}/\expval{O_1} \approx 0.9521$, while the numerical result gives $n_{s}/\expval{O_1} \approx 0.9710$.
From Eq.~\eqref{eq:AC_cond_simple}, the the superfluid density can be read as $n_{s}/\expval{O_1} = 1$.
Indeed, Eq.~\eqref{sigmaAC1} can be said as a better approximation than Eq.~\eqref{eq:AC_cond_simple} but the correction is very small and difficult to see in Fig.~\ref{fig:conductivity_low_T}.
Note that Fig.~\ref{fig:conductivity_low_T} corresponds to the solution above $T=k/(2\sqrt{2}\pi)$ so the solution with the dilatonic black hole is not favored actually.
We could not obtain the superconducting solution in $T<k/(2\sqrt{2}\pi)$, where the dilatonic black hole is the ground state, for the $O_1$ theory.

\subsection{Low temperature approximation}
Now, we show another approximation of the AC conductivity in the superconducting phase following the method studied in \cite{Siopsis:2010uq}.
Unfortunately, however, it does not provide useful results in our model.

In superconducting phase, to study the AC conductivity, we need to solve Eq.~\eqref{Axz}. 
Imposing the infalling boundary condition at the horizon, we write
\begin{equation}\label{Az}
	A_{x}(z)=\left(1-z\right)^{\frac{-i\omega}{4\pi T}}G(z)
\end{equation}
where $G(z)$ is a regular function at $z=1$.
$G(z)$ can be expanded as
\begin{equation}
	G(z)=\alpha_{0}+\alpha_{1}(1-z)+\alpha_{2}(1-z)^{2}+\cdots \ .
\end{equation}
Substituting Eq.~\eqref{Az} into Eq.~\eqref{Axz}, we get
\begin{multline}\label{G(z)}
	G^{\prime\prime}(z)+\left(\frac{2}{z}+\frac{i\omega}{2\pi T(1-z)}+\frac{f^{\prime}(z)}{f(z)}+\frac{\phi^{\prime}(z)}{\sqrt{3}}\right)G^{\prime}(z)+\\
	\frac{1}{48z^{4}}\Biggl\{-\frac{3\omega z^{3}\left(4i\pi T\left(-2+z\right)+\omega z\right)}{\pi^{2}T^{2}\left(-1+z\right)^{2}}+\\
	\frac{4}{f(z)^{2}}\left(12\omega^{2}r_{h}^{2}-24e^{-\frac{\phi(z)}{\sqrt{3}}}q^{2}r_{h}^{2}f(z)\Phi(z)^{2}-\frac{i\omega z^{4}f(z)\left(3f^{\prime}(z)+\sqrt{3}f(z)\phi^{\prime}(z)\right)}{\pi T\left(-1+z\right)}\right)\Biggr\}G(z)=0
\end{multline}
The equation involves $\Phi(z)$.
We focus on the $O_1$ theory here.

In order to solve Eq.~\eqref{G(z)}, we utilize the method which develops in \cite{Siopsis:2010uq}.
We rewrite the coordinate as $z = \varepsilon\zeta$, and expand the equation for small $\varepsilon$.
We choose $\varepsilon = 2\pi T/\expval{O_1}$ to consider the low temperature limit.
For small $\varepsilon$, Eq.~\eqref{G(z)} becomes
\begin{equation}
    \pdv[2]{G}{\zeta}
    -\left(1-\tilde{P}\right)G
    + \varepsilon\frac{i\tilde{\omega}}{\sqrt{1-\tilde{P}}} \pdv{G}{\zeta} + \order{\varepsilon^2} = 0,
\end{equation}
where $\tilde{P} = P/r_h$ and $\tilde{\omega} = \omega/r_h$.
We truncate this equation up $\order{\varepsilon^2}$.
The general solution is obtained as
\begin{equation}
    G =
    e^{
    -\frac{i\omega/r_h}{\sqrt{1-P/r_h}}\varepsilon\zeta
    }
    \left\{
    c_1 e^{-\zeta\sqrt{
        \left(1-\frac{P}{r_h}\right)
        - \frac{\varepsilon^2}{4} \frac{\omega^2/r_h^2}{1-P/r_h}
    }}
    +
    c_2 e^{\zeta\sqrt{
        \left(1-\frac{P}{r_h}\right)
        - \frac{\varepsilon^2}{4} \frac{\omega^2/r_h^2}{1-P/r_h}
    }}
    \right\}.
\end{equation}
In the original $z$ coordinate, we write
\begin{equation}\label{Gz}
    G(z)\approx
    e^{
    -\frac{i\omega/r_h}{\sqrt{1-P/r_h}} z
    }
    \left\{
    c_1 e^{+\frac{z \varepsilon^{-1}}{\sqrt{1-P/r_h}}}
    +
    c_2 e^{-\frac{z \varepsilon^{-1}}{\sqrt{1-P/r_h}}}
    \right\}.
\end{equation}
Here, we have dropped the higher $\varepsilon$ terms in the exponential terms.
Recalled $A_x(z)= (1-z)^{\frac{-i\omega}{4\pi T}}G(z)$, the conductivity is obtained as
\begin{equation}\label{sigma}
    \sigma(\omega)
    \approx
    i\frac{\expval{O_1}}{\omega}
    \frac{1-c_1/c_2}{1+c_1/c_2}.
\end{equation}
Now, we consider fixing the ratio $c_1/c_2$ by using the equation at $z=1$:
\begin{equation}\label{G[1]}
    G'(1) \left(1-\frac{i \tilde{\omega}}{\sqrt{1-\tilde{P}}}\right)
    +\frac{1}{16} G(1) \left(16 q^2 \Phi(1)^2
    - 4 \tilde{\omega}^2 \frac{1-2\tilde{P}}{(1-\tilde{P})^2}
    - \frac{4 i \tilde{\omega}}{\sqrt{1-\tilde{P}}}
    \right) =0.
\end{equation}
Note that the equation involves unknown parameter $\Phi(1)$.
We assume that the scalar profile can be approximated by $\Phi(z)=\frac{\expval{O_1}}{r_h \sqrt{2}} z$.
Then, $\Phi(1)$ is given by $\Phi(1)=\expval{O_1}/(r_h\sqrt{2})$.
Using Eqs.~\eqref{Az}, \eqref{Gz} and \eqref{G[1]}, the ratio $c_{1}/c_{2}$ becomes ($\omega\rightarrow 0$)
\begin{equation}
    \frac{c_{1}}{c_{2}} \approx
    e^{2\frac{\sqrt{1-\tilde{P}}}{\varepsilon}}\left[
        1 + \frac{
            2 \tilde{O}_1^2(1-\tilde{P})^2 - i \tilde{\omega} \left(
                3 (1-\tilde{P})^{3/2} - i \tilde{\omega}(3 - 4\tilde{P})
            \right)
        }{2(1-\tilde{P})^2(-i\tilde{\omega} + \sqrt{1-\tilde{P}})} \varepsilon
    \right],
\end{equation}
where $\tilde{O}_1 = \expval{O_1}/r_h$.
We have dropped the higher $\varepsilon$ terms again.
Plugging this to \eqref{sigma}, the conductivity is written as
\begin{equation}
\sigma(\omega) \approx
\frac{i \tilde{O}_1}{\tilde{\omega}}\left[
    1 - 2 e^{-2 \tilde{O}_1} \left(
        1 + \frac{
            2 \tilde{O}_1^2(1-\tilde{P})^2 - i \tilde{\omega}(
                3(1-\tilde{P})^{3/2} - i \tilde{\omega}(3 - 4 \tilde{P})
            )
        }{2 \tilde{O}_1 (1 - \tilde{P})^{3/2} (\sqrt{1-\tilde{P}} - i \tilde{\omega})}
    \right)^{-1}
\right].
\end{equation}
The low frequency expansion is obtained
\begin{equation}
    \sigma(\omega) \approx
    \frac{i\tilde{O}_1}{\tilde{\omega}}\left[
        1 - \frac{2}{1+\tilde{O}_1} e^{-2\tilde{O}_1}
        + i \tilde{\omega} \frac{2 \tilde{O}_1^2 - 3}{\tilde{O}_1 (\tilde{O}_1 + 1)^2 \sqrt{1-\tilde{P}}} e^{-2\tilde{O}_1}
        + \order{\tilde{\omega}}
    \right].
\end{equation}
The leading coefficient corresponds to the superfluid density which can be read
\begin{equation}\label{ns2}
    n_s \approx \expval{O_1} - \frac{2 \expval{O_1}}{\expval{O_1} + r_h} e^{-2 \expval{O_1}/r_h}.
\end{equation}
However, the approximation becomes better when $\expval{O_1}\gg r_h$, so the contribution from the second term is very small.
Neglecting the second term, we obtain the same result for $n_s$ read from Eq.~\eqref{eq:AC_cond_simple}.


\end{document}